\documentclass[twocolumn,twocolappendix]{aastex63}
\usepackage[utf8]{inputenc}
\usepackage{placeins}
\usepackage{graphicx}
\usepackage{amsmath}

\newcommand{\inv}{$^{-1}$ }
\newcommand{\invnospace}{$^{-1}$}

\begin{document}

\title{The Debris of the ``Last Major Merger'' is Dynamically Young}

\author[0000-0002-7746-8993]{Thomas Donlon II}
\affiliation{Department of Physics, Applied Physics and Astronomy, Rensselaer Polytechnic Institute, 110 8th St, Troy, NY 12180, USA}
\affiliation{Department of Physics and Astronomy, University of Alabama in Huntsville, 301 North Sparkman Drive, Huntsville, AL 35816, USA}
\correspondingauthor{Thomas Donlon II}
\email{thomas.donlon@uah.edu}

\author[0000-0001-8348-0983]{Heidi Jo Newberg}
\affiliation{Department of Physics, Applied Physics and Astronomy, Rensselaer Polytechnic Institute, 110 8th St, Troy, NY 12180, USA}

\author[0000-0003-3939-3297]{Robyn Sanderson}
\affiliation{Department of Physics \& Astronomy, University of Pennsylvania, 209 S 33rd St., Philadelphia, PA 19104, USA}
\affiliation{Center for Computational Astrophysics, Flatiron Institute, 162 5th Ave., New York, NY 10010, USA}

\author[0000-0003-3792-8665]{Emily Bregou}
\affiliation{Department of Physics \& Astronomy, University of Pennsylvania, 209 S 33rd St., Philadelphia, PA 19104, USA}

\author[0000-0003-1856-2151]{Danny Horta}
\affiliation{Center for Computational Astrophysics, Flatiron Institute, 162 5th Ave., New York, NY 10010, USA}
\affiliation{Astrophysics Research Institute, Brownlow Hill, Liverpool, L3 5RF, UK}

\author[0000-0002-8354-7356]{Arpit Arora}
\affiliation{Department of Physics \& Astronomy, University of Pennsylvania, 209 S 33rd St., Philadelphia, PA 19104, USA}

\author[0000-0001-5214-8822]{Nondh Panithanpaisal}
\affiliation{Department of Physics \& Astronomy, University of Pennsylvania, 209 S 33rd St., Philadelphia, PA 19104, USA}

\begin{abstract}
The Milky Way's (MW) inner stellar halo contains an [Fe/H]-rich component with highly eccentric orbits, often referred to as the ``last major merger.'' Hypotheses for the origin of this component include \textit{Gaia}-Sausage/Enceladus (GSE), where the progenitor collided with the MW proto-disk 8-11 Gyr ago, and the Virgo Radial Merger (VRM), where the progenitor collided with the MW disk within the last 3 Gyr. These two scenarios make different predictions about observable structure in local phase space, because the morphology of debris depends on how long it has had to phase mix. The recently-identified phase-space folds in \textit{Gaia} DR3 have positive caustic velocities, making them fundamentally different than the phase-mixed chevrons found in simulations at late times. Roughly 20\% of the stars in the prograde local stellar halo are associated with the observed caustics. Based on a simple phase-mixing model, the observed number of caustics are consistent with a merger that occurred 1--2 Gyr ago. We also compare the observed phase-space distribution to FIRE-2 Latte simulations of GSE-like mergers, using a quantitative measurement of phase mixing (2D causticality). The observed local phase-space distribution best matches the simulated data 1--2 Gyr after collision, and certainly not later than 3 Gyr. This is further evidence that the progenitor of the ``last major merger'' did not collide with the MW proto-disk at early times, as is thought for the GSE, but instead collided with the MW disk within the last few Gyr, consistent with the body of work surrounding the VRM. \\\vspace{0.5cm}
\end{abstract}


\section{Introduction} \label{sec:intro}

The Milky Way's (MW) stellar halo was built up through accretion events \citep{Helmi2020}. There is substantial evidence that the MW's stellar halo is predominantly composed of the debris from a single massive progenitor dwarf galaxy \citep{Deason2015,Deason2018,Myeong2018,Belokurov2018b,Vincenzo2019,IorioBelokurov2019,Mackereth2019,Feuillet2020,Naidu2020,Han2022a}, which is thought to have been accreted early on in the MW's history \citep{Deason2013,Gallart2019,Naidu2021}. This merger event is commonly referred to as \textit{Gaia}-Sausage/Enceladus \citep[GSE,][]{Belokurov2018,Helmi2018}, and it is often considered to be the MW's ``last major merger.''

There are several cross-consistent arguments for why the GSE merger occurred 8-11 Gyr ago. One point is that we observe a smooth, monotonically decreasing broken power law density profile in the MW stellar halo \citep{Yanny2000,Juric2008,Sesar2013,Xue2015,Hernitschek2018,Han2022a}. Recent accretion events that have not had time to sufficiently relax will have non-monotonic density profiles. \cite{Deason2013} argued that this indicates that the merger that is responsible for the break in the MW stellar halo density had to be at least 4.5 Gyr old, based on mixing timescales from the \cite{BullockJohnston2005} simulations. 

Further dynamical evidence for the GSE being accreted at early times comes from the observation that the GSE debris is on high-eccentricity orbits \citep{Belokurov2018}. Halo star orbits radialize over time due to the mass growth of the host galaxy; it makes sense that the debris of an ancient merger event would have had a long time to radialize, which is why the GSE stars are on high-eccentricity orbits at the present day.

Another argument for the ancient age of the GSE comes from the ages of GSE and thick disk stars. One possible explanation for why the MW has both a thick disk and a thin disk is that the MW proto-disk was heated by a collision with a massive object; this elevated the thick disk stars to orbits with large vertical heights, at which point gas could re-condense to form the thin disk \citep{QuinnGoodman1986,VelazquezWhite1999,Grand2020}. Because the thick disk cannot continue forming stars after being heated, the time of collision corresponds to the ages of the youngest thick disk stars ($\gtrsim$8 Gyr), which agrees with the observed ages of GSE stars \citep{Schuster2012,Helmi2018,Gallart2019}. 

Further, the observed chemical bimodality in MW disk stars \citep{Bensby2014} can potentially be explained by an accretion of metal-poor gas onto the MW disk at early times, which could be caused by a gas-rich dwarf galaxy colliding with the disk \citep{Chiappini1997,Helmi2018,Buck2020}. Additionally, an in-situ halo population (sometimes called the ``Splash'') has been discovered in the MW's inner stellar halo, which is characterized by [Fe/H]-rich stars on orbits with large vertical height and high eccentricity \citep{Bonaca2017,Belokurov2020}. This population is also thought to have been heated from stars in the MW's proto-disk, probably from the same merger that created the thick disk.


However, recent work has called into question whether a massive, ancient merger event is actually required to explain the observed properties of the MW system. It has been shown that simulations can create both chemical and kinematic bimodalities in galaxy disks without any merger events \citep{Clarke2019,BeraldoeSilva2020}, and there is evidence for thin disk stars that are older than the proposed collision time of the GSE \citep{BeraldoeSilva2021}. The in-situ stellar halo can also be generated via secular processes that do not involve a collision with another galaxy at early times, such as lumps in the proto-disk \citep{Amarante2020,BelokurovKravtsov2022,Rix2022,Conroy2022} and/or stars that are formed in outflows and fall back into the inner halo \citep{Yu2020}. Recent analysis of cosmological simulations also indicates that massive radial mergers occur at late times with some regularity \citep{Horta2023}.

The dynamics of the GSE debris also seem to be at odds with the GSE progenitor dwarf galaxy colliding with the MW 8-11 Gyr ago. \cite{Donlon2019} were able to recreate the observed GSE halo substructure using a simulation of a dwarf galaxy colliding with the MW only 2 Gyr ago; they claimed that longer integration times did not accurately reproduce the observed structure. Additionally, \cite{Donlon2020} identified stellar shells in the MW, and compared the observed shell structure to simulations of radial mergers; they found that the MW's stellar shells appeared to be caused by a dwarf galaxy colliding with the MW 2.7 Gyr ago, and if the collision had happened more than 5 Gyr ago then the observed shell structure would be too phase-mixed to resolve individual shells. This is consistent with earlier assessments that the debris of satellites is well-mixed within a a few Gyr after collision with the host galaxy disk \citep[e.g.][]{VillalobosHelmi2008}. However, \cite{Donlon2020} used tailored $N$-body simulations in a static potential to evaluate the phase-mixing rates of radial mergers; in cosmological settings, it is possible that interactions with other merger events or the time-dependent potential of the host galaxy could alter the rate of phase mixing for radial merger debris.

Because the shells were associated with a halo structure called the Virgo Overdensity \citep{Vivas2001,Newberg2007,Juric2008,Bonaca2012,Donlon2019}, and were not associated with an ancient merger, the shell structure and associated debris was named the Virgo Radial Merger (VRM). The VRM has since been shown to have distinct chemistry and dynamics compared to other structures that are typically included all together in the set of GSE stars; \cite{Donlon2022} and \cite{Donlon2023} showed that the MW experienced many radial mergers over a range of times. In particular, the progenitor of the [Fe/H]-rich debris that forms shells in the inner halo most likely collided with the MW disk around 3 Gyr ago, rather than 8-11 Gyr ago. The structure known as GSE is actually a combination of several distinct accretion events, each of which has a unique chemodynamic distribution. 



Recently, \cite{Belokurov2023} presented the phase-space folds (variably called chevrons or caustics) of the GSE debris in the \textit{Gaia} DR3 data \citep{GaiaDR3,GaiaDR3RVS}. They showed that the phase-space folds were present in stars with [Fe/H] $<$ -0.7, which is consistent with GSE debris. Once a dwarf galaxy collides with its host, it dissolves and its stars are free to phase mix in the host galaxy potential; as a result, the number of folds and their physical properties provide us with a diagnostic tool to measure how long the stars in those structures have been unbound from their progenitor galaxy \citep{HernquistQuinn1989,Donlon2020}. 

In this work, we show that a simple semi-analytical model of phase mixing indicates that the observed number of caustics in the MW are consistent with a recent collision time. We compare the phase-space folds in the local Solar neighborhood to a MW-mass system containing a GSE-like merger in a cosmological zoom-in simulation, in order to determine whether the observed local halo substructure is consistent with an ancient massive radial merger. We find that the phase-space caustics indicate that the radial debris typically associated with the GSE is consistent with a dwarf galaxy that collided with the MW disk within the last 1--2 Gyr, and is not consistent with a collision 8-11 Gyr ago. These results are generally consistent with those of \cite{Donlon2020}, indicating that in a cosmological setting, radial mergers are expected to phase mix at roughly the same rate, or perhaps slightly faster, than similar tailored $N$-body simulations.

\subsection{Terminology}

In this work we follow the terminology laid out by \cite{Donlon2020}, and expand upon it when necessary. Here is a list of relevant definitions from a top-down perspective: a radial merger event consists of a dwarf galaxy on a radial orbit becoming unbound and depositing its constituent stars, or merger debris, into the stellar halo of its host galaxy; the debris of radial mergers form caustics, or caustic structures, which describe the morphology of the radial merger debris in $v_r-r$ phase space; stars with $v_r\sim0$ located in caustics form shells, which are apocenter pileups for stars within some energy range in a given caustic. 

We also use the terms ``chevron'' and ``phase space fold'' in this work. ``Phase space fold'' is a catch-all term used to describe observable substructure in $v_r-r$ phase space; while they are often caused due to the presence of unmixed radial merger debris in the data, it is possible that phase space folds can originate from other phenomena. Here we use ``chevron'' only to refer to phase space folds that are phase-mixed, and therefore are not caustics; we elaborate the dynamical differences between chevrons and caustics in later sections. While we treat caustics and chevrons as quantitatively different types of structures in this work, note that some literature uses the two terms interchangeably.

\section{A Semi-Analytical Caustic Timing Model} \label{sec:semi-analytical}

\begin{figure}%
\centering
\includegraphics[width=0.45\textwidth]{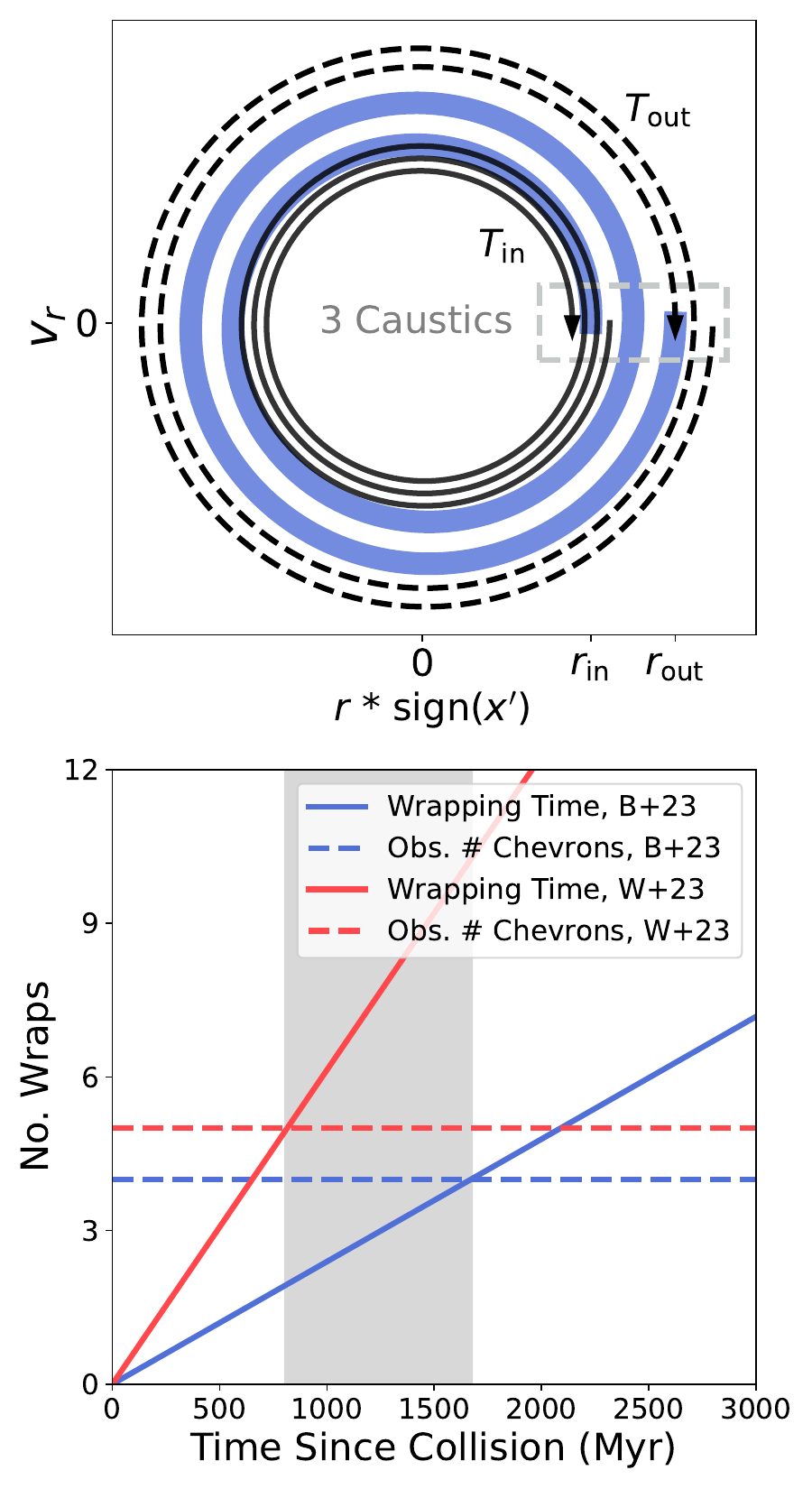}
\caption{A simple semi-analytical model for estimating the time of a radial collision given the number of observed caustics in a range of the Galaxy. The top panel illustrates phase mixing in $r$-$v_r$ phase space; an initial Gaussian perturbation to the phase space density between $r_\textrm{in} < r < r_\textrm{out}$ will wrap over a timescale given by the orbital frequencies of the innermost and outermost material. A new caustic is created when the innermost material ``laps'' the outermost material; as a result, the number of caustics on a side of the Galaxy can be related to the collision time of the merger that generated the observed caustics. The horizontal axis is a deprojected radial coordinate, equal to $r$ multiplied by the sign of a Galactocentric Cartesian coordinate oriented so that the merger event falls in along the $x'$ axis (avoiding compression at $r=0$). The bottom panel shows the rate that new caustics are generated given observed caustic data in the MW from \cite{Belokurov2023} (B+23, blue) and \cite{Wu2023} (W23, red). The collision times inferred by these studies indicate that the progenitor of the phase space folds collided with the MW disk roughly 1--2 Gyr ago. }\label{fig:semi-analytical}
\end{figure}

The number of caustics in a given region of space that arise from the debris of a single merger event increases over time as the debris phase mixes \citep{Donlon2020}. Here, we derive a simple relation for the number of caustics within a given distance range as a function of time; we can then estimate the time of a merger event given the observed number of caustics and the extent of their apocenter radii. 

Before experiencing disruption, the tidal debris of a radial merger will populate a localized overdensity in $r$-$v_r$ phase space. After the dwarf has been disrupted during passage through the Galactic center, the merger debris behaves like a collection of test particles in the host galaxy's gravitational potential. The orbital frequency of the material depends on each star's distance from the Galactic center; stars that are initially closer to the center of the Galaxy will have a shorter orbital period than stars that begin farther away from the center of the Galaxy. The merger debris will then begin to wrap in phase space (similar to the $z$-$v_z$ phase space wrapping observed in \citealt{Antoja2018}). The top panel of Figure \ref{fig:semi-analytical} shows this physical process.

The number of shells observed on one side of the host galaxy increases by 1 every time the innermost material ``laps'' the outermost material in phase space. The angle between the innermost material and outermost material in phase space is \begin{equation}
    \theta_\textrm{in} - \theta_\textrm{out} = 2\pi t \left(\frac{1}{T_\textrm{in}} - \frac{1}{T_\textrm{out}}\right),
\end{equation} where $T_\textrm{in}$ and $T_\textrm{out}$ are the azimuthal orbital periods of the innermost and outermost material, respectively. The time it takes for the debris to wrap once, i.e. form another shell, is then when $\theta_\textrm{in} - \theta_\textrm{out} = 2\pi$; \begin{equation}
    t_\textrm{wrap} = \frac{T_\textrm{out}T_\textrm{in}}{T_\textrm{out} - T_\textrm{in}},
\end{equation} and the number of caustics within $r_\textrm{in} < r < r_\textrm{out}$ is \begin{equation}
    n_\textrm{wrap} - 1 = \frac{t}{t_\textrm{wrap}}.
\end{equation} Here, $n_\textrm{wrap} - 1$ is used because there will be two caustics after a single wrapping time. Note that the total number of shells in the entire host galaxy can be computed as $n_\textrm{shells} = 2 n_\textrm{wrap}$, because we must count shells on both sides of the host galaxy. 

In order to estimate the collision time of a given radial merger, we need (i) the number of caustics within $r_\textrm{in} < r < r_\textrm{out}$, and (ii) the orbital periods for orbits with apocenters at $r_\textrm{in}$ and $r_\textrm{out}$. For the first point, \cite{Belokurov2023} claim that there are 5 distinct caustics with apocenters between 11.5 kpc and 25 kpc; \cite{Wu2023} claim that there are 6 caustics with apocenters between 6.75 kpc and 30.25 kpc. For the second point, \cite{BinneyTremaine2008} give a rough estimate of the time it takes material on a radial orbit with a given apocenter $r_\textrm{ap}$ to oscillate from one side of the Galaxy to the other and back; \begin{equation}
    2\;T(r_\textrm{ap}) = 4\int_0^{r_\textrm{ap}} \frac{\textrm{d}r}{\sqrt{2\left[\Phi(r_\textrm{ap}) - \Phi(r)\right]}},
\end{equation} which we integrate numerically using the \textit{Gala MilkyWayPotential2022} \citep{Gala} model for the underlying gravitational potential. An additional factor of 2 is required to account for the fact that the particle will be located on the other side of the Galaxy after a single radial orbital period.

This produces $t_\textrm{wrap} = $ 420 Myr for the \cite{Belokurov2023} data, and $t_\textrm{wrap} = $ 160 Myr for the \cite{Wu2023} data (see the bottom panel of Figure \ref{fig:semi-analytical}). These data estimate the collision time of the progenitor of the observed MW caustics to be 1680 Myr and 800 Myr, respectively, indicating that the current data are consistent with a collision time within the last 1--2 Gyr ago. This estimated collision time is somewhat more recent than previous estimates of the merger time of the VRM \citep[$\sim$3 Gyr ago, ][]{Donlon2020}; however, these values are well outside the estimated time of the GSE collision, suggesting that the observed caustics are generated by a much more recent merger event. The various caveats and implications of the semi-analytical model are discussed further in Section \ref{sec:discussion}. 

In order to explore a more realistic determination of the collision time that considers various effects of a radial merger in a time-dependent gravitational potential, we will compare the observed MW data to that of a cosmological $N$-body simulation. While the cosmological simulation reveals interesting dynamical features that are not present in this simple model, the relevant dynamics of the systems are ultimately similar, and as a result we obtain a nearly identical estimate of the collision time of the MW's last major merger. 

\begin{figure*}[!ht]%
\centering
\includegraphics[width=0.85\textwidth]{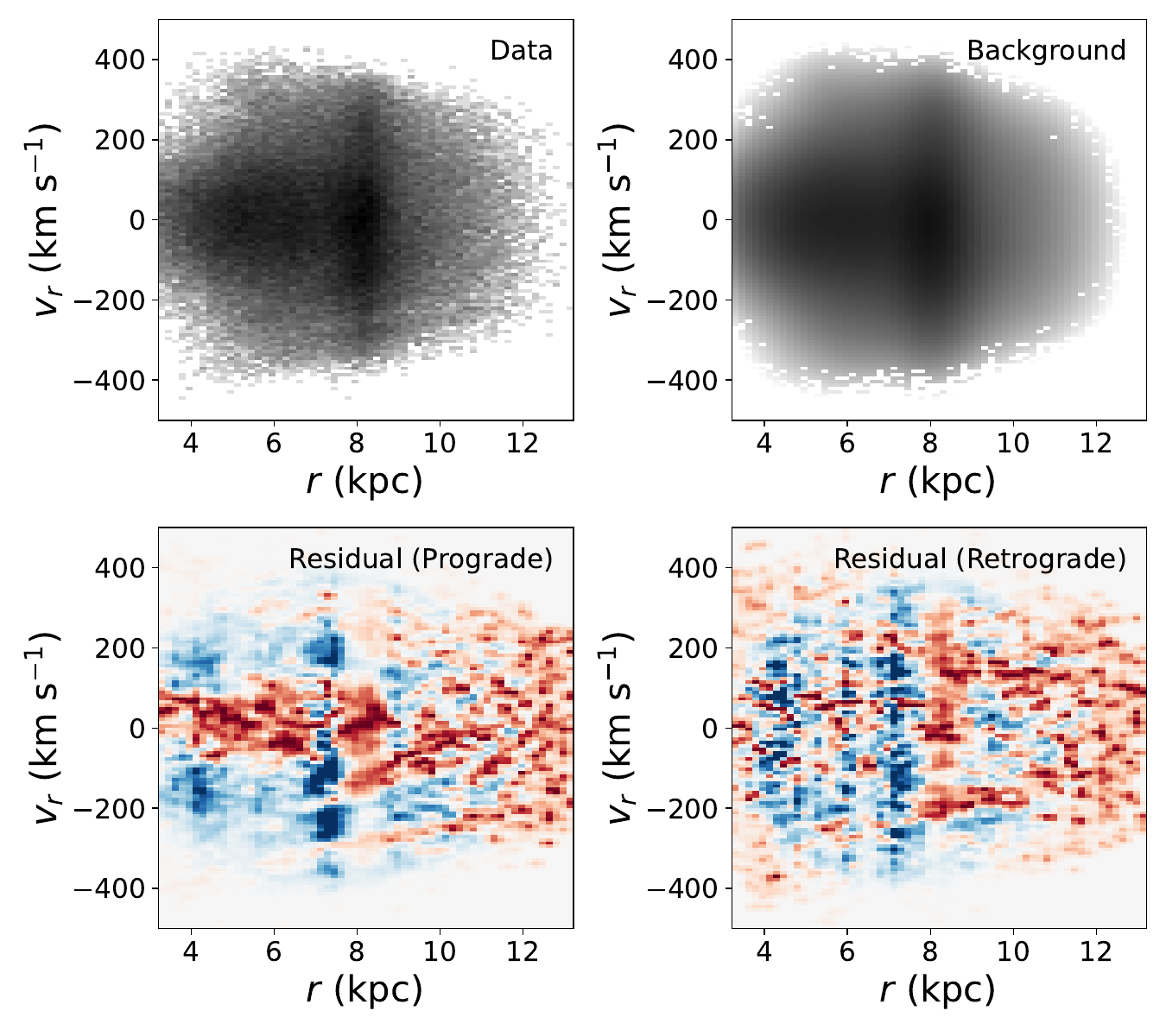}
\caption{Sample of local stars from \textit{Gaia} DR3 with high quality phase-space measurements and $|L_z|<500$ kpc kms\invnospace. The top left panel shows the observed distribution of the data in $r$-$v_r$ phase space; the phase-space folds are barely visible as faint wrinkles in the density of stars, particularly in stars with $v_r<0$. The dark vertical band at the Solar radius is due to higher sampling efficiency near the Sun. The top right panel shows a background model constructed from a low-pass filter on the observed data, which should not contain small-scale phase-space substructure. Subtracting this background model from the observed data produces a residual composed of small-scale structure; the residual for stars with prograde and retrograde angular momenta are shown in the bottom panels, which have been column normalized. These residuals show alternating red and blue diagonal stripes (red indicates an excess in density, blue shows a depletion), which correspond to the phase-space folds of the local stellar halo. The residual in the prograde stars appears to be asymmetric about the $v_r=0$ line, whereas the retrograde material contains a single large chevron, which is symmetric about the $v_r = 0$ line.}\label{fig:data}
\end{figure*}

\section{Data Selection} \label{sec:data}

Here we collect the data to reproduce the observed local phase space folds. We selected all stars from the \textit{Gaia} DR3 RVS catalog that satisfied the following criteria: \begin{itemize}
    \item Relative parallax uncertainty less than 10\%, and positive parallax (before offset correction),
    \item Relative proper motion and radial velocity uncertainties less than 20\%,
    \item \verb!RUWE! $<$ 1.25 \citep[to eliminate potential binary stars,][]{Penoyre2022},
    \item Located within 5 kpc from the Sun, and
    \item Not located within 1.5$^\circ$ of \cite{Harris1996} globular clusters within 5 kpc of the Sun.
\end{itemize}

The distance to each star was computed from its inverse parallax, after correcting for the parallax zero-point offset using the \verb!gaiadr3_zeropoint! package \citep{Lindegren2021}. We used a Solar position of $R_\odot$ = 8.17 kpc from the Galactic center \citep{GravityCollaboration2019} and $Z_\odot$ = 21 pc above the disk plane \citep{BennettBovy2019}, and a Solar reflex motion of $v_\odot$ = (9.3, 251.5, 8.59) km s\inv \citep{GaiaCollaboration2022}. 

We also enforced that each star had $|L_z| < 500$ kpc km s\inv, which removes most of the nearby disk stars and restricts our sample to mostly GSE debris \citep{Buder2022,Belokurov2023}. After these cuts, 132,934 stars remained in the dataset.

\section{The Local Phase-Space Caustics}

Figure \ref{fig:data} shows the observed \textit{Gaia} data. The majority of stars in the data are not expected to be members of the phase space folds, so there will be a substantial amount of background that needs to be removed in order to clearly see the local halo substructure; however, the phase space folds are barely visible as wrinkles in the density of stars in the top left panel of Figure \ref{fig:data} even before background subtraction, particularly in stars with $v_r<0$. 

We constructed a background by first binning the observed data distribution into 60 bins in $r$ and 120 bins in $v_r$ (each bin is 0.167 kpc by 4.167 km s\invnospace). We then convolved the observed binned distribution with a 2-dimensional Gaussian with a standard deviation of 5 pixels in $r$ and 15 pixels in $v_r$; this will preserve only the structure on scales larger than the Gaussian kernel from the observed data. This background is shown in the top right panel of Figure \ref{fig:data}.

The background was then subtracted from the observed data distribution in order to extract the small-scale structure from the observed data. This was done separately for stars on prograde and retrograde orbits, because \cite{Belokurov2023} show that the observed phase space folds have different morphology in stars with opposite signs of angular momenta; the residuals for these data are shown in the bottom row of Figure \ref{fig:data}. 

Both residuals contain apparent alternating blue and red diagonal stripes, which corresponds to the phase-space folds identified by \cite{Belokurov2023}. The prograde residual contains several blue and red alternating lines, which are asymmetric about the $v_r=0$ line. In contrast, the retrograde residual contains only a single large chevron, which appears to be symmetric about the $v_r = 0$ line. The importance of the different symmetry in these residuals will be emphasized later on in this work; for now it is sufficient to note that the morphology of the prograde structures and the retrograde structure are not the same.

\subsection{Number of Stars in the Phase Space Caustics}

We can estimate the fraction of the local stellar halo that is in these phase space caustics by comparing the number of stars in the data with the sum of the absolute magnitude of the residual histogram, \begin{equation} \label{eq:n_stars}
    N_{*,resid} = \sum_i^{n_b} |n_{i,resid}|,
\end{equation} where $n_b$ is the number of bins, and $n_{i,resid}$ is the number of stars in a given bin in the residual. The ratio of $N_{*,resid}$ to the total number of stars in the data corresponds to the strength of the residual signal, and therefore the fraction of stars that are present in the underlying caustic structures. The prograde data contains 91,994 stars and $N_{*,resid}$ = 17,596, so the stars in the prograde phase space caustics make up 19\% of the stars in the prograde local stellar halo. The retrograde material produces a similar result, with 40,940 stars in the dataset and $N_{*,resid}$ = 10,094, so the stars in the retrograde phase space caustics contribute 25\% of the stars in the retrograde local stellar halo. Combining the prograde and retrograde data results in an estimated contribution of 21\% of the local stellar halo stars by the stars in these caustics.

If the stars in the phase space caustics belong to the GSE debris, as has been suggested by \cite{Belokurov2023}, then this estimate is much lower than those of \cite{Belokurov2018} and \cite{Helmi2018}, who claimed that the GSE merger event contributes the ``bulk'' (at least half) of the stars in the MW stellar halo. However, this estimate is similar to the recent estimate of \cite{Lane2023} that the GSE debris could contribute as little as 10-20\% of the local stellar halo stars. Our estimate here is also slightly lower than the estimated fraction of local halo stars contributed by the VRM, which is supposed to contribute $\sim35\%$ of the local halo stars \citep{Donlon2023}.

\section{Ancient Merger Simulation}

\begin{figure*}%
\centering
\includegraphics[width=0.85\textwidth]{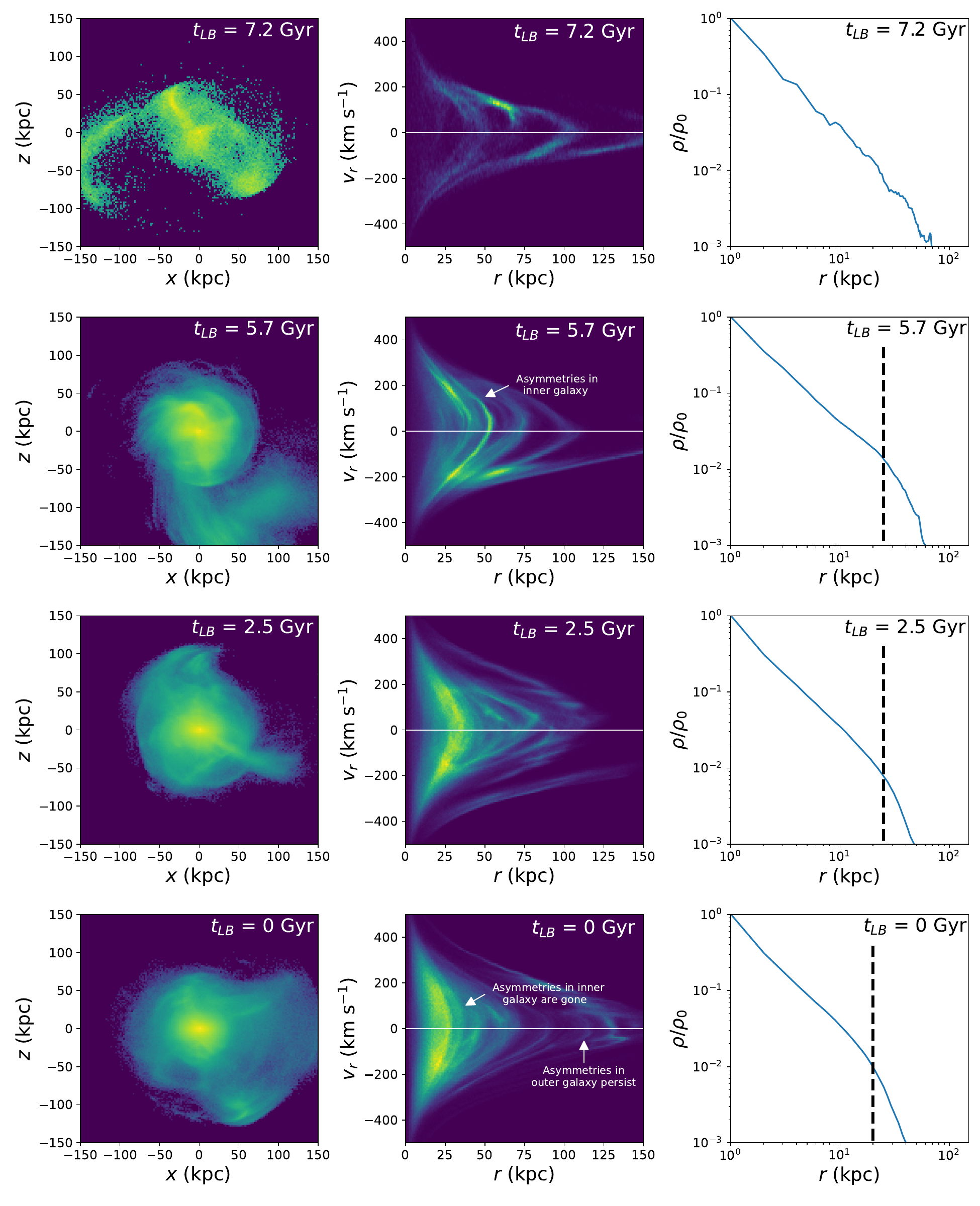}
\caption{Upsampled simulated ancient radial merger event from the m12f Latte FIRE-2 cosmological zoom-in simulation. Each row contains the data for a given snapshot, with the top row earliest in the simulation progressing forwards in time to the bottom row, which is at the present day. The left column shows the $X$-$Z$ configuration space distribution of the merger debris, the middle column shows the $r$-$v_r$ phase space distribution of the merger debris, and the right panel shows the global density profile of the merger debris. The progenitor dwarf galaxy of the merger debris becomes unbound at a lookback time of 8.8 Gyr, and collides with the host galaxy disk at a lookback time of 8.5 Gyr. Note that even at early times, the global density profile appears to monotonically decrease. The phase space distribution of the merger debris becomes more symmetric about $v_r=0$ (solid white line in the middle column) over time; this process occurs more quickly in the inner galaxy ($r\lesssim 30$ kpc) than in the outer galaxy ($r\gtrsim 30$ kpc). At late times, asymmetric complex structure and wrinkles can still be seen in the caustics in the outer galaxy, while the inner galaxy debris forms large chevrons which are symmetric about $v_r=0$. This suggests that the level of asymmetry about $v_r=0$ in the inner galaxy is a good assessment of how phase-mixed merger debris is (how long it has been since its progenitor became unbound).}\label{fig:simulation}
\end{figure*}

Our goal is to determine the time that has passed since the progenitor of the local phase-space folds collided with the MW disk. While we have an estimate of this time from a simple dynamical model, it is beneficial to compare the observed data with a simulation in which we can pinpoint the exact time that a dwarf galaxy collides with its host galaxy. The simulated data can then be used to study the phase-space morphology of the merger debris as a function of time since the collision, and verify that the additional interactions and time-dependent potential in the cosmological simulations do not substantially change the timing result that we get from the simple dynamical model.

\cite{Donlon2020} used a series of simulations of a single dwarf galaxy falling into a static MW potential in order to quantify how radial mergers phase mix over time. However, a simulation in a static potential does not include the effects of dynamical friction, time-dependent potentials (including bars and other halo substructure), and the reaction of the disk, which are all expected to influence the evolution of merger debris. Recent works have used radial mergers from cosmological simulations in order to include more of these dynamical processes, thus improving the quality of the simulated models \citep{Belokurov2023,Davies2023a}. 

In this work, we analyze two of these simulated cosmological merger events. The first merger event occurs early on in its host galaxy's history, and is chosen to be an analog of a GSE merger. This ancient merger event is the focus of this section. The second merger event occurs late in its host galaxy's history, and is chosen to be an analog for a recent radial merger such as the VRM; Section \ref{sec:recent_rme} discusses the recent merger event in detail. 

Since any simulated merger event will be heavily dependent on the conditions of the simulation, we urge the reader to not think of this study as statistical proof that all early radial mergers behave identically. Rather, we use this simulation as an indication that the dynamics of cosmological radial mergers are not wildly different than expected. We also use these the simulated data to explain why other studies have inferred a large time since collision for the same phase space fold data. Ultimately, we find that the timing results of the cosmological simulation are consistent with those derived from the simple model in Section \ref{sec:semi-analytical}.

\subsection{Choice of Merger}

We selected our simulated data from the Latte suite of FIRE-2 cosmological zoom-in simulations \citep{Wetzel2016,Hopkins2018}. These simulated galaxies are chosen as a representative set of MW-like galaxies in that they form compact stellar disks and have total mass similar to that of the MW, although each simulation has a unique accretion history. Additionally, previous works have constructed catalogs of all major merger events in the Latte halos \citep{Samuel2020,Panithanpaisal2021,Shipp2022,Santistevan2023,Horta2023}, as well as their properties, which provides a foundation for analyzing the dynamics of specific phase-mixed structures in these simulations. 

The ancient merger event used in this work was selected as our GSE analog because it satisfied three properties: (i) The progenitor dwarf galaxy collided with its host galaxy 8-11 Gyr ago, (ii) the progenitor dwarf galaxy is sizeable compared to the mass of the disk at the time of collision, and (iii) the progenitor dwarf galaxy falls in on a high-eccentricity orbit, so it forms stellar shells and its constituent stars are on radial orbits at $z$=0. We found three suitable candidates in the \cite{Horta2023} catalog of Latte simulation merger events (one each in m12f, m12i, and m12w); we selected the merger event in m12f because it appeared the most similar overall to the predicted properties of the GSE. 

One could argue that an ancient progenitor of the GSE debris could have become unbound at early times while on a low-eccentricity orbit, and that the GSE debris is radially anisotropic today because the mass growth of the MW radialized the orbits of those stars over time. However, it is difficult to see how this process would form shells or caustics, which have been observed in the MW, so we choose to use a radial merger event for our GSE analog in order to match the observed properties of the GSE debris.

\begin{figure}%
\centering
\includegraphics[width=0.45\textwidth]{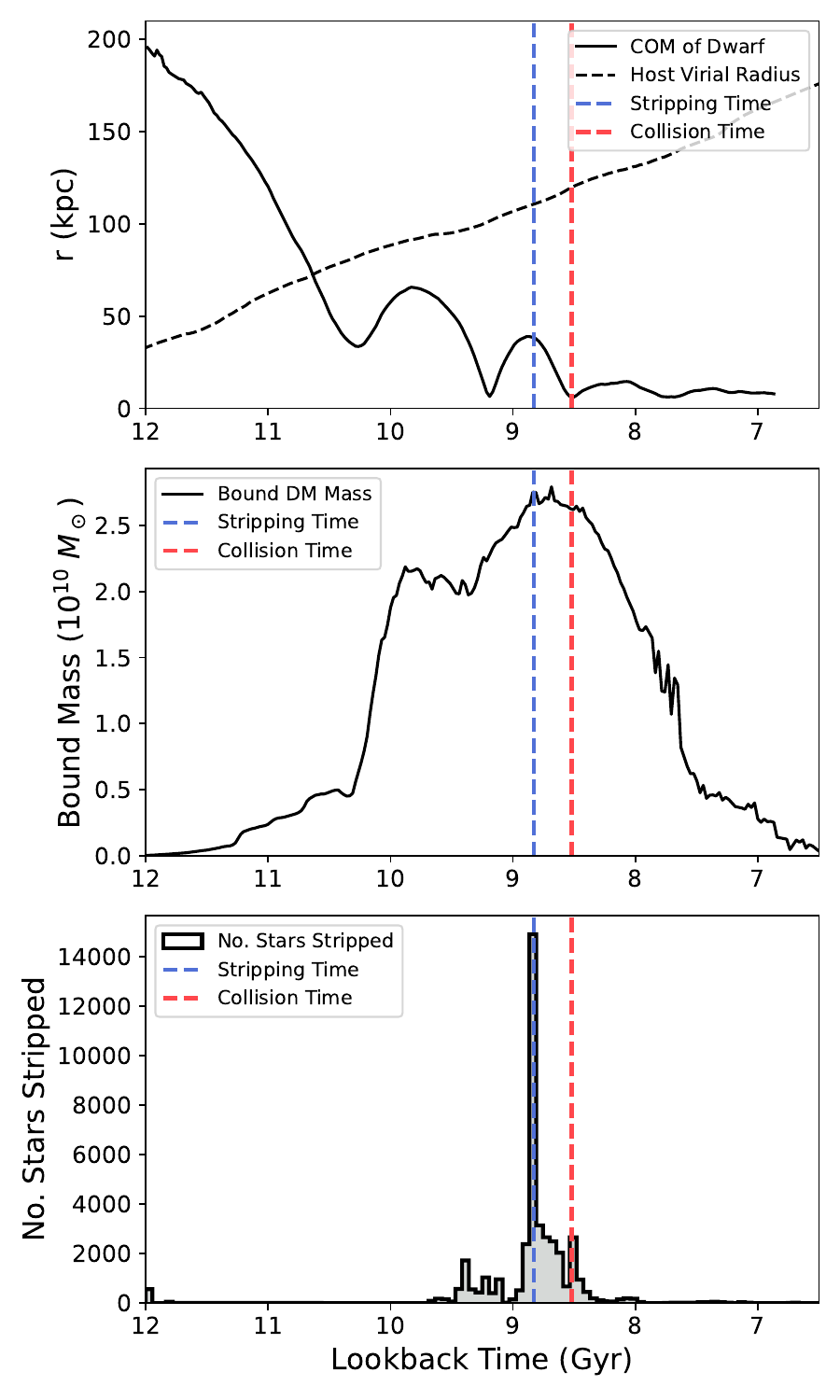}
\caption{Position and bound mass of the progenitor dwarf galaxy for the first few Gyr of the simulation. \textit{Top}: The position of the center of mass of the progenitor with respect to the host galaxy disk (solid black line). The virial radius of the host galaxy is given as a dashed black line. The dashed blue and red lines correspond to the time when the majority of stellar material is stripped from the progenitor, and the time when the progenitor collides with the host disk, respectively. The collision occurs at the progenitor's third pericenter passage. \textit{Middle:} The mass of dark matter bound to the dwarf galaxy as a function of lookback time. There appears to be a minor dark matter stripping event (without corresponding stripping of baryonic matter) after the first pericenter passage. After collision, the bound mass rapidly decreases. \textit{Bottom:} The number of stars stripped from the progenitor as a function of lookback time. Nearly all stars are stripped at a lookback time of 8.8 Gyr, while the progenitor was at its last apocenter before collision.}\label{fig:infall}
\end{figure}

Several snapshots of the simulated ancient merger event are shown in Figure \ref{fig:simulation}, and the orbital path and bound mass of its progenitor dwarf galaxy are shown in Figure \ref{fig:infall}. The progenitor dwarf galaxy of this merger event has a dark matter mass of 2.8$\times10^{10} M_\odot$ and a stellar mass of $2.0\times10^8M_\odot$ at its final pericenter passage; the host galaxy disk has a stellar mass of $2.1\times10^{10}M_\odot$ at the time of collision, making the dwarf galaxy nearly 50\% more massive than the host disk at the time of collision. The majority of its constituent stars become unbound from the dwarf galaxy at a lookback time of 8.8 Gyr ago, and the center of mass of the progenitor dwarf galaxy passes through the host galaxy disk 0.3 Gyr later. The final apocenter radius of the simulated progenitor dwarf galaxy's orbit is 39 kpc, which is not very different from the predicted final apocenter radius of the last major merger \citep{Deason2013,Han2022a}. 

The metallicity distribution of the progenitor dwarf galaxy peaks at [Fe/H] = -1.5, which is slightly below the observed metallicity of GSE stars ([Fe/H] = -1.2, \citealt{Naidu2020}); this discrepancy is explained by the observation that FIRE dwarfs have systematically lower metallicities than their observed MW counterparts by roughly 0.3 dex at masses consistent with the GSE \citep{Panithanpaisal2021}.

\subsection{Upsampling}

In order to meaningfully compare the simulated data to the observed data, the simulated particles must be dense enough in phase space that our residual contains several particles per bin, where the bins are in $v_r$ and $r$ as described in Section \ref{sec:data}. Unfortunately, the simulation data is too sparse, so it needed to be upsampled in order to obtain the resolution necessary for analysis. 

In Section \ref{sec:data}, we estimated the number of stars in the local phase space caustics to be roughly 27,000 stars. Because it is still possible to see the caustics in the data after splitting the data into prograde and retrograde halves, we choose 27,000/2 = 13,500 stars to be the desired number of stars in a mock local solar neighborhood for our simulation to adequately resolve caustics. 

We took the simulated merger data at the each snapshot and computed how many particles were located on average in a mock Solar region. Each mock Solar region consisted of a 5 kpc sphere centered on a point 8 kpc from the center of the host galaxy. We used 8 mock Solar regions, each rotated $\pi/4$ radians from each other with respect to the center of the host galaxy. At the early snapshots where upsampling was possible, the average mock Solar region contained $\sim$100 particles, which is 0.7\% of our desired number of particles for a mock solar region. In order for all snapshots to have the desired resolution, we would need to upsample by a factor of 135; however, we decided to upsample by a factor of 100 rather than 135 because (i) over 90\% of the simulation contained enough particles in the average mock solar region where 100x upsampling would provide adequate resolution, and (ii) the 100x upsample takes nearly one month to run on available computing resources, and we wanted to limit the amount of computation required to complete the upsampling process.

The upsampling procedure consisted of determining the phase-space coordinates of all particles belonging to the progenitor of the chosen radial merger event at a lookback time of 7.3 Gyr, and replacing them with 100 points distributed in a Gaussian with a spatial dispersion of 6 pc and velocity dispersion of 5 km/s, which is roughly the scale of an open cluster \citep{Sanchez2009,Soubiran2018}. We then integrated all of these test particles forwards in the time-dependent interpolated potential model described in \cite{Arora2022,AroraInPrep}; this allowed us to realistically obtain the orbits of these new particles without needing to re-run the entire simulation. 

We chose to upsample the particles at a lookback time of 7.3 Gyr because (i) it was after all the particles had been stripped from the progenitor dwarf galaxy, so they were effectively test particles that could be treated without self-gravity, and (ii) the time-dependent potential model only went back to a lookback time of 7.3 Gyr.

The mean phase space distance between each particle and its nearest neighbor in the simulation at a lookback time of 7.3 Gyr is 500 pc in space and 2 km s\inv in velocity; the scale chosen for the upsampling Gaussian kernel results in a mean phase space distance between each particle and its nearest neighbor of only 2 pc in space and 1 km s\invnospace in velocity. By upsampling the data on a scale smaller than the characteristic scale of the simulation, we have underestimated the amount of phase mixing that has occurred up until a lookback time of 7.3 Gyr (and therefore we underestimate the amount of phase mixing at all times in the upsampled simulation). Therefore, our estimates of phase mixing and collision times in the upsampled simulation are probably biased high.

\subsection{Numerical Resolution \& Artificial Stripping}

It has been pointed out that simulated dark matter halos are often artificially disrupted due to low resolution \citep{vandenBoschOgiya2018}. It is important to evaluate whether we have sufficient resolution in our simulation in order to properly model this specific merger event. \cite{vandenBoschOgiya2018} provide two criteria for evaluating whether artificial disruption substantially affects a simulated dark matter halo; the first tests whether the progenitor is subject to low force resolution, and the second tests whether the progenitor is subject to discreteness noise. Both criteria are given as inequalities in \cite{vandenBoschOgiya2018} that we evaluated using the properties of the progenitor dwarf galaxy at each snapshot in the not upsampled simulation; the simulated dwarf galaxy is subject to artificial stripping when the inequality is no longer satisfied.

We find that the simulated dwarf galaxy maintains the required resolution to avoid artificial stripping even after the primary stripping event at a lookback time of 8.8 Gyr. The resolution of the simulation does not become problematic until the dwarf galaxy has reached a fraction of a percent of its peak bound mass, at which point the gravity of the dwarf remnant is not expected to strongly affect the orbits of the stripped stars.

Additionally, Appendix A of \cite{Barry2023} indicates that increased force and mass resolution in the FIRE simulations did not substantially change the subhalo number density, which suggests that the resolution of the FIRE simulations is sufficient for examining properties of the halo.

\section{Evaluations of Phase Mixing}

\subsection{Radial Density Profile}

One argument for the early accretion of the GSE progenitor is that recent accretion events were expected to have poorly-defined radial density profiles that were not monotonically decreasing; they therefore did not look like the observed smooth component of the stellar halo. \cite{Deason2013} argued that accreting satellites did not mix quickly enough to have well-defined density profiles until $\sim$4.5 Gyr after becoming unbound.

The time evolution of the simulated ancient merger's radial density profile is shown in the right column of Figure \ref{fig:simulation}. The top row of Figure \ref{fig:simulation} shows that the merger debris has assembled a well-defined global density profile by 1.5 Gyr after the unbinding time, which was at a lookback time of 8.8 Gyr. This indicates that phase mixing occurs in radial mergers on much shorter timescales than presented by \cite{Deason2013}, possibly because they were using semi-analytical simulations with static potential models \citep{BullockJohnston2005}. Another possibility is that the satellites in those simulations had orbits with a variety of eccentricities, and satellites on circular orbits will have longer mixing times (although it should be noted that satellites on circular orbits are not good analogs for the GSE merger). 

Within 3 Gyr after becoming unbound (the second row of Figure \ref{fig:simulation}), the simulated merger debris appears to be fairly uniform in configuration space, excluding some tidal features at large radii. We argue that the rapid assembly of a well-defined density profile in the simulated data indicates that the observed density profile of the MW inner stellar halo does not rule out a recent merger event.

\subsection{Causticality} \label{sec:causticality}

\begin{figure*}%
\centering
\includegraphics[width=0.95\textwidth]{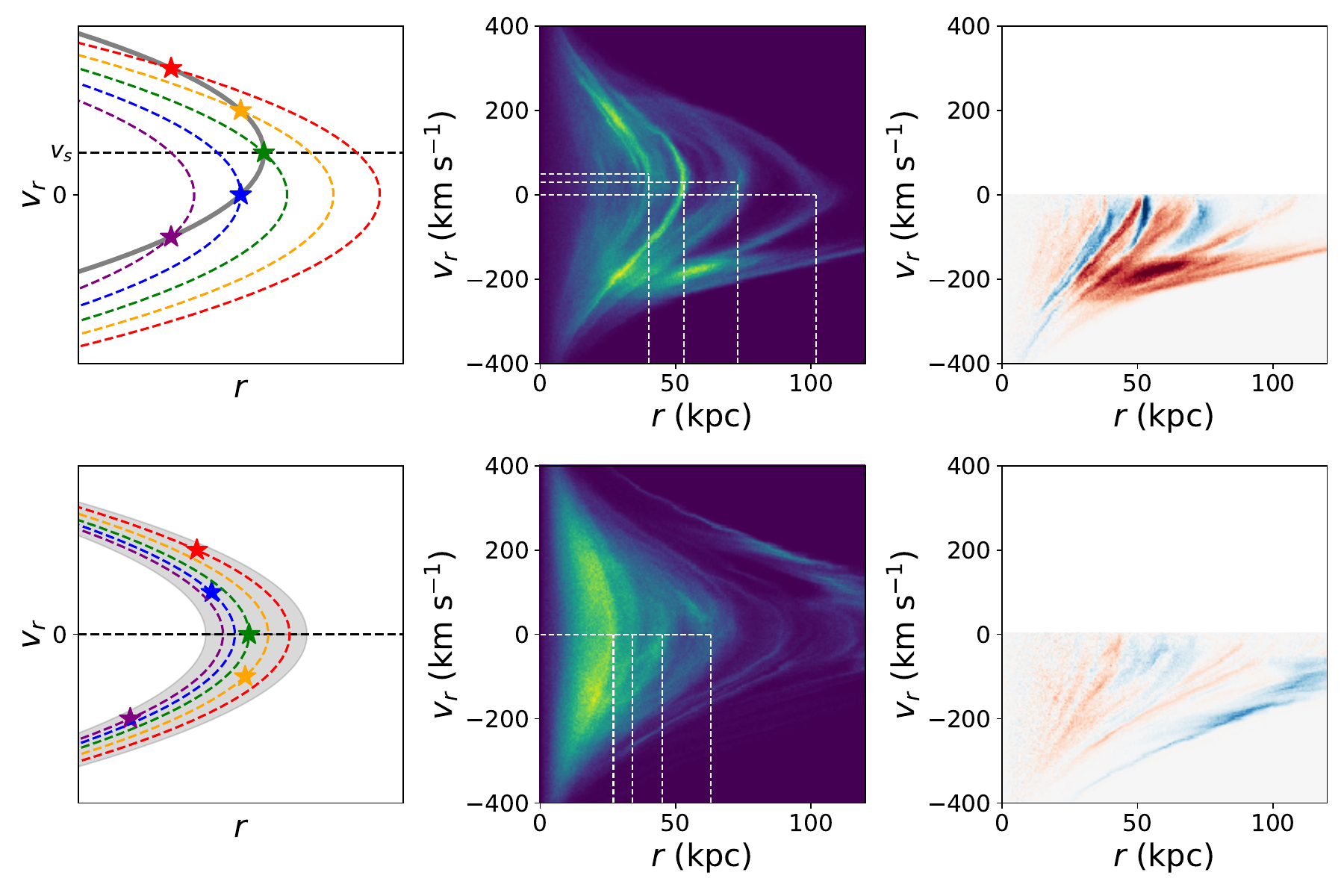}
\caption{An explanation of how phase mixing, shell velocity ($v_s$), and causticality are related. \textit{Left:} Orbits of individual stars (colored stars and colored dashed lines) compared to a phase-space fold (solid gray line) with a given $v_s$ (dashed black line). The top panel shows how a caustic with $v_s>0$ is constructed of energy-sorted stars (highest energy is red, lowest energy is purple), where orbital energy increases with $v_r$. The bottom panel shows a phase-mixed chevron, where the individual stars are not energy-sorted (higher $v_r$ stars don't necessarily have larger energy), so they form a thick chevron (gray region) rather than a narrow caustic. \textit{Middle:} Upsampled simulated data, with white dashed lines marking $r_s$ and $v_s$ for various phase space folds. The top panel shows the simulated data at a lookback time of 5.7 Gyr, which is not very phase-mixed; as a result, its caustics have $v_s>0$, and $v_s$ for each caustic decreases as $r$ increases. The bottom panel shows the simulated data at the present day, which is very phase-mixed; its phase-space folds all have $v_s=0$, and are much thicker than the caustics in the top panel. \textit{Right:} The folded residuals from the middle panel, scaled to the same color range. The folded residual for the earlier snapshot has a larger magnitude in most bins because it is very asymmetric about $v_r=0$ (it has a large causticality). The folded residual for the present-day snapshot has smaller magnitude in each bin because it is more symmetric about $v_r=0$ (it has small causticality). The top middle panel is identical to the middle panel in the second row of Figure \ref{fig:simulation}, but zoomed in.}\label{fig:phase_mixing}
\end{figure*}

We noted in Figure \ref{fig:simulation} that the amount of symmetry about the $v_r=0$ line is a good proxy for how phase mixed the merger debris is. Here we compare the local distribution of stars in phase space with the time series of our cosmological ancient merger simulation, with a quantitative assessment of this symmetry. 

\cite{Donlon2020} defined a quantitative metric, called ``causticality,'' which measures the amount of phase mixing on a scale from 1 (not at all phase-mixed) to 0 (completely phase-mixed). Their definition used 1-dimensional histograms in $r$, because they only had information on the radii of stellar shells in the MW halo; they were unable to identify entire caustics in $v_r-r$ phase space using the available data at the time. Because this work deals with phase mixing in a 2-dimensional slice of phase space, we must extend the definition of causticality to 2 dimensions to make it most useful for comparing our data to the simulation.

We define 2-dimensional causticality as \begin{equation} \label{eq:c}
    C = \frac{\sum_i^{n_{\textrm{bins}}} (n_{i,v^+} - n_{i,v^-})^2}{\sum_i^{n_{\textrm{bins}}} (n_{i,v^+} + n_{i,v^-})^2}, 
\end{equation} where the notation is the same as in Equation \ref{eq:n_stars} except $n_{i,v^+}$ and $n_{i,v^-}$ are the corresponding bins of the phase space data with positive and negative $v_r$. This procedure can be intuited as folding the phase space distribution across the $v_r=0$ line and then computing a normalized residual sum of squares between the two halves of the phase space distribution; as a result, 2-dimensional causticality is a measure of the symmetry of a given phase-space distribution about the $v_r=0$ line. The uncertainty in this value is derived in Appendix \ref{app:sigma_c}. This definiton of causticality is still bounded as $0\leq C \leq 1$. $C=1$ when no stars in a structure have a counterpart when reflected across $v_r=0$ (for example, if all stars have positive $v_r$); $C=0$ when the phase space distribution of a structure is perfectly symmetric about $v_r=0$. 

\begin{figure}%
\centering
\includegraphics[width=0.45\textwidth]{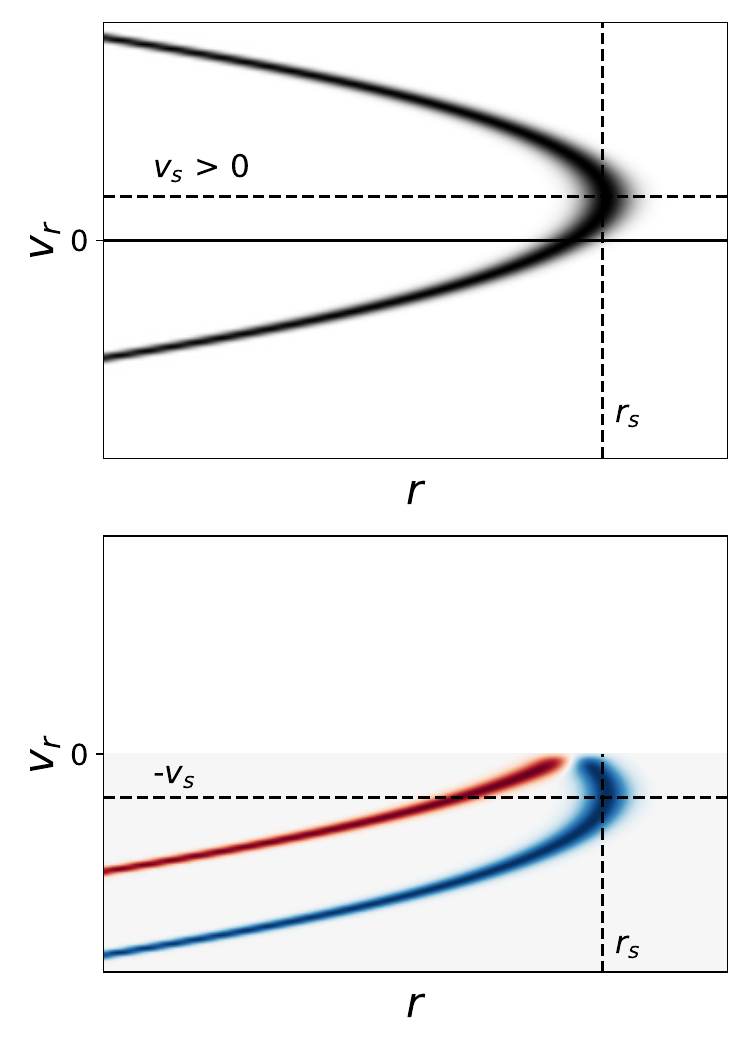}
\caption{The folded residual for a single model caustic. \textit{Top:} An analytical phase-space density model for a single tidal caustic using the model from  \cite{SandersonHelmi2013}. The caustic is approximately parabolic, with a turn-around point at $r_s$ and $v_s>0$; this is a consequence of a caustic being built up from many energy-sorted stars with different orbits, rather than being a single orbital path. \textit{Bottom:} Folded residual of the analytical caustic in the top panel. Red corresponds to more density with $v_r<0$, blue corresponds to more density with $v_r>0$. The blue and red regions are offset from one another because $v_s\neq0$; if $v_s=0$ for this structure, then the blue and red regions would perfectly overlap and cancel in the folded residual. \label{fig:chevrons}}
\end{figure}

Figure \ref{fig:phase_mixing} shows why causticality works as a metric for phase mixing; a caustic has a positive non-zero ``shell velocity'' at the point where its constituent stars currently have maximum $r$ \citep{SandersonHelmi2013}, which makes it reflection asymmetric about $v_r=0$. Initially in radial merger events, there are only a few caustics in the merger debris (e.g. the top row of Figure \ref{fig:phase_mixing}), which have nonzero $v_s>0$, which makes the phase-space distribution asymmetric about $v_r=0$. The asymmetry of a single caustic is shown in Figure \ref{fig:chevrons}. As these caustics mix, more and more shells are created \citep{HernquistQuinn1989,Donlon2020}. Eventually, caustics begin to pile up on top of other caustics, at which point the phase space distribution becomes a single thick chevron (e.g. the bottom row of Figure \ref{fig:phase_mixing}). This single thick chevron is not energy-sorted because it is not a single caustic, so it appears to have a turn-around point at $v_s=0$, making it symmetric about $v_r=0$. 

One quirk about this definition is that the presence of any (symmetric) background will dramatically reduce the measured causticality, because the denominator of Equation \ref{eq:c} becomes very large compared to the numerator. This is not a problem for the simulated data, which has no background. The observed data has a background, but it can be removed following the example in Figure \ref{fig:data}. As long as there is no background, $C$ does not depend on the number of stars/particles present in the data, so we do not need to worry about normalizing the observed data to the simulated data (however, the lower noise threshold of causticality depends on the particle resolution, as shown in Appendix \ref{app:saturation}). However, the choice of binning scheme can alter $C$ for the same dataset, so we require that the observed data and simulated data are identically binned. 

\begin{figure*}[!ht]%
\centering
\includegraphics[width=\textwidth]{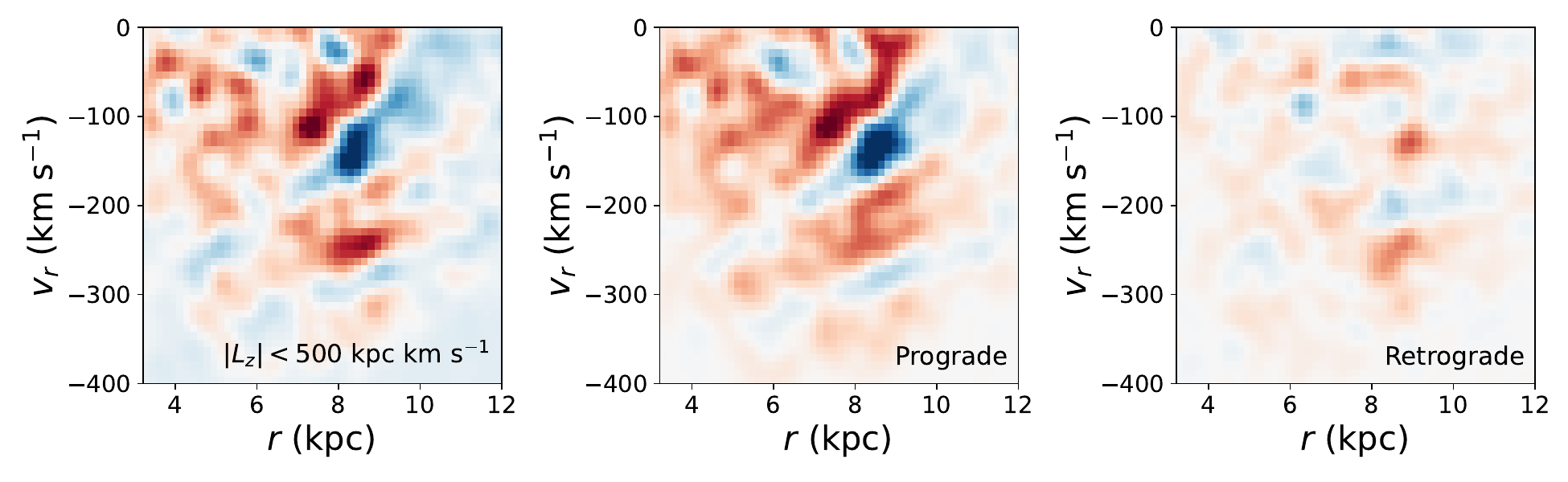}
\caption{The observed local phase space split up by angular momentum. Each panel has been convolved with a Gaussian filter to reduce noise. Colors are the same as in Figure \ref{fig:chevrons}, and the color scale is identical for all panels. The left panel contains all stars with $|L_z|<500$ kpc km s\invnospace, whereas the middle and right panels show only the stars from the left panel with prograde and retrograde $L_z$, respectively. The prograde stars prominently display alternating red and blue bands, which indicate caustics, while the retrograde stars lack any coherent structure. The chevron in the bottom right panel of Figure \ref{fig:data} is not present in the folded residual for the retrograde stars, suggesting that it is an old structure. This is in contrast to the results of \cite{Belokurov2023}, who claimed that the prograde and retrograde material in the local stellar halo contain different phase-space folds from the same merger event.}\label{fig:lz}
\end{figure*}

\subsection{Causticality of Observed Data}

Folded residuals for the observed data after background subtraction are shown in Figure \ref{fig:lz}. The left panel of this figure contains all of the observed data after cuts, and four caustics are apparent as sets of red and blue alternating bands. As before, we split up the data into its prograde and retrograde stars, which are known to contain different substructure. 

Interestingly, all of the prominent caustic structures are contributed by the prograde stars, which appear nearly identical to the folded residuals of the prograde + retrograde stars. The chevron in the bottom right panel of Figure \ref{fig:data} does not appear in the retrograde data folded residual due to its symmetry about $v_r=0$; this suggests that (i) the chevron in the retrograde data is dynamically old because it has $v_s=0$, and as a result, (ii) the caustics in the prograde data, which appear to come from a more recent merger event, and the chevron in the retrograde data, which appears to come from an older merger event, probably arise from different merger events. This is inconsistent with the analyses of \cite{Belokurov2023} and \cite{Davies2023b}, who claimed that the local phase-space folds all arise from the same merger event. Rather, this result agrees with the findings of \cite{Donlon2022} and \cite{Donlon2023}, who claimed that the local stellar halo contains debris from many radial mergers with a variety of accretion times; this could easily explain why we appear to see phase space folds with different dynamical ages in the local phase space.

We choose to use only the prograde stars to measure the causticality of the observed data, because the folded residual of the prograde stars contains clear caustic structures, whereas the folded residual of the retrograde stars appears to be consistent with random noise. The causticality of this data is $C$ = 0.5188 $\pm$ 0.0007. Adding the retrograde and prograde stars together and then calculating the causticality of the observed data produces $C$ = 0.4250 $\pm$ 0.0004, although adopting this lower value for the causticality of the observed data does not meaningfully change the conclusions of this work. Additionally, the observed causticality of the data depends on the choice of bin size (see Appendix \ref{app:binning}), although because the observed and simulated data use the same binning scheme, no systematic error is introduced by the choice of bin size.

\subsection{Causticality of Simulated Data}

\begin{figure*}%
\centering
\includegraphics[width=\textwidth]{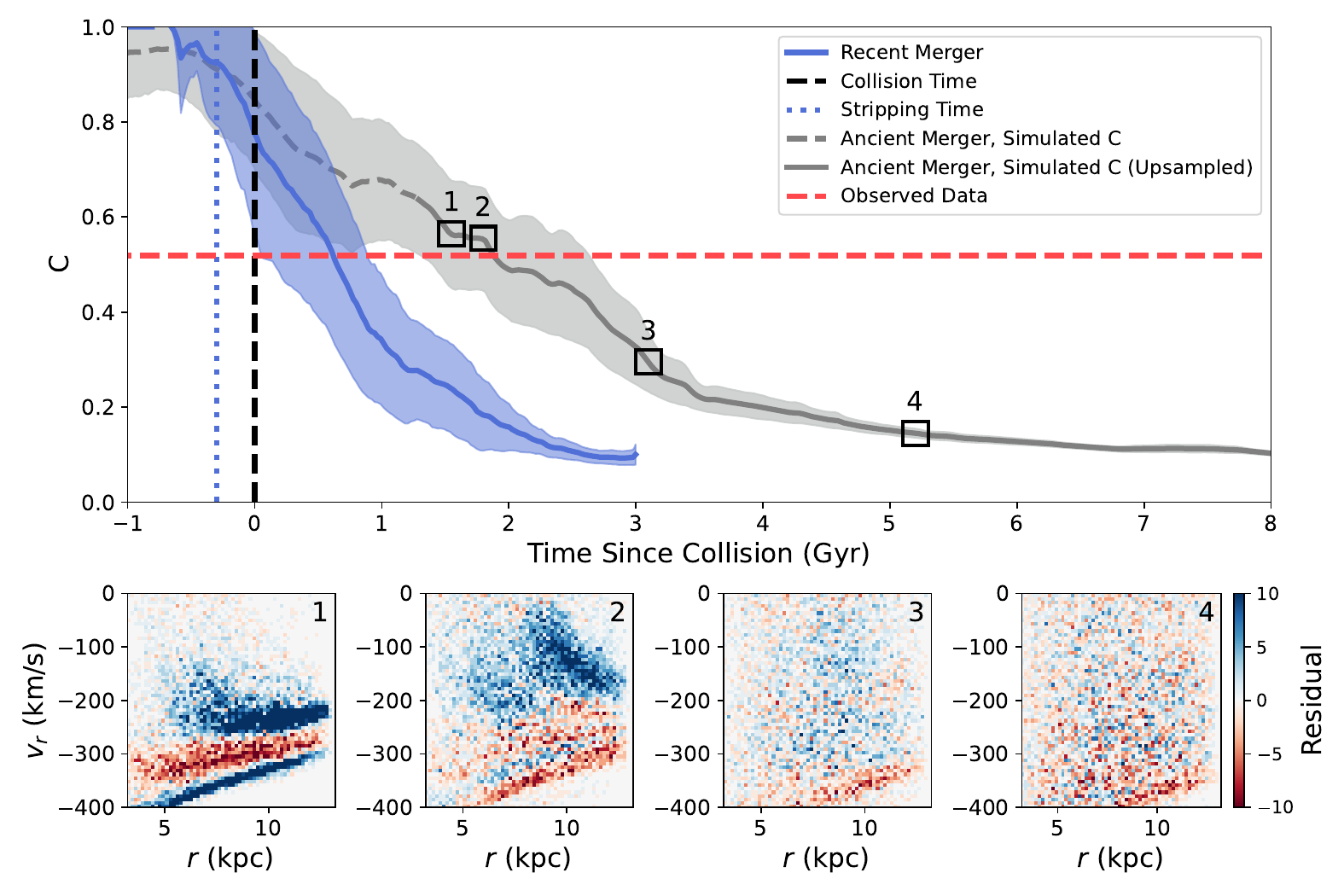}
\caption{Mean causticality of the simulated ancient merger (grey line) as a function of time. The 1$\sigma$ uncertainty in this value is shown as a light gray region. The dashed blue and black lines indicate the time at which the progenitor dwarf galaxy became unbound and collided with the host galaxy disk, respectively. The dashed red line indicates the causticality of the observed data. The causticality of the simulated data steadily decreases over time as the merger debris phase mixes; the simulated data and the observed data have similar causticality for a few Gyr after the progenitor dwarf collides with the host galaxy, but have very different causticality after $\sim4$ Gyr after collision. This indicates that the progenitor of the local phase space folds must have collided with the MW disk within the last $\sim$3 Gyr. The dashed gray line at early times corresponds to the causticality of the merger debris data that could not be upsampled due to not having time-dependent potential models for those snapshots. The causticality of the simulated recent merger event is shown in blue; its causticality drops much more quickly than that of the ancient merger event. The causticality of the observed data is consistent with that of the recent merger debris much earlier than the ancient merger debris -- as recently as $\sim$0.6 Gyr ago. The recent merger event simulation abruptly cuts off at 3 Gyr since collision because its simulation ends at that time. The bottom panels show examples of the folded residuals of simulated caustics from the recent merger simulation; the numbers of the panels indicate the time and causticality of each bottom panel in relation to the top panel.}\label{fig:causticality}
\end{figure*}

To compare the causticality of the simulated data with that of the observed data, we computed $C$ for the 8 mock Solar regions in the simulation, for all 300 simulation snapshots. Causticality as a function of time is shown in Figure \ref{fig:causticality}, compared to the causticality of the observed data. 1$\sigma$ uncertainty for the simulated data are shown as a shaded gray region; this uncertainty is computed as the standard deviation of the causticality values measured from all 8 mock solar regions, although the individual uncertainties in causticality for each solar region are much smaller than this shaded region. This provides an idea of the uncertainties in causticality that are generated from noise and the choice of location within the Galaxy from which we are measuring (as the amount of phase-mixing in each mock solar region is expected to be similar, but is not necessarily exactly the same at a given time). 

The observed and simulated $C$ measurements only overlap for the first $\sim3$ Gyr after the progenitor dwarf galaxy collides with the host galaxy disk. The observed and simulated $C$ curves intersect roughly 2 Gyr after collision; this is one estimate for the time of collision between the progenitor of the local caustics and the MW disk. 

This is somewhat smaller than the estimated collision time from \cite{Donlon2020}. The discrepancy is probably due to the fact that a static potential damps phase mixing due to back-reactions of the host galaxy on the merger debris \citep{Seguin1994,Seguin1996}; the live galaxies in the FIRE-2 simulations are more realistic MW models for studying phase mixing than an analytical potential model. This result is inconsistent with the local phase space folds being caused by a collision that occurred 8-11 Gyr ago, with a significance of nearly 28$\sigma$.

\section{Recent Merger Simulation} \label{sec:recent_rme}

\subsection{Phase Mixing Rates At Different Times} \label{sec:phase_mixing_rates}

The rate at which the debris of a radial merger event phase mixes will depend on the shape and strength of the gravitational potential of its host galaxy. Because a galaxy's potential is expected to be substantially different today than it was at high redshift, our simulation of a dwarf galaxy colliding with its host galaxy's disk at early times may not accurately represent how quickly the phase-mixing would have occurred if that progenitor had instead collided with its host galaxy 3 Gyr ago. As a result, we caution the reader not to over-interpret the time when the observed and simulated causticality intersect in Figure \ref{fig:causticality} as a measurement of when the progenitor of the local caustics collided with the MW disk (especially given the large uncertainty in the simulated causticality at early times).

The rate of phase-mixing can be thought of as being directly proportional to the orbital frequency of the debris. Because galaxies grow in mass over time, an orbit at late times with a given radius will have a shorter orbital period (and therefore more rapid phase-mixing) than an orbit with the same radius at early times. This means that phase-mixing of a major radial merger event will only occur more quickly within the last several Gyr compared to our simulation, and that we expect that the time of collision estimated in Figure \ref{fig:causticality} between the MW disk and the progenitor of the local caustics is biased towards later times.

In the following sections, we test these ideas against a recent radial merger event from the same set of simulations as our ancient simulated merger event, and explore the differences in the values of causticality measured for these structures as a function of time.

\;\\

\subsection{Recent Radial Mergers in the Latte Simulations}

\begin{figure}%
\centering
\includegraphics[width=0.5\textwidth]{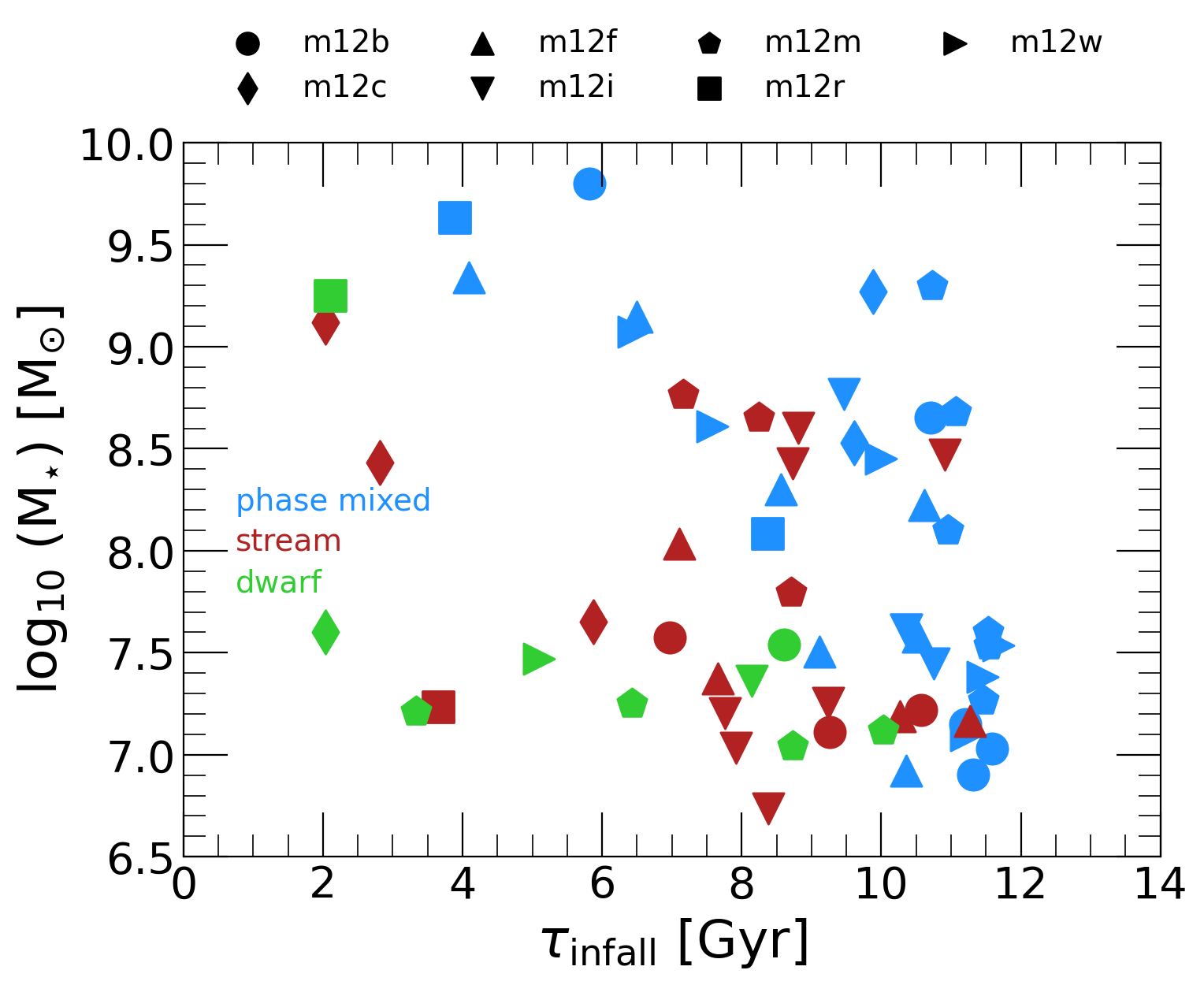}
\caption{Stellar mass of each major accretion event in the Latte simulations at the time when the progenitor crossed the virial radius of its host galaxy, vs the lookback time of virial crossing. The shape of each point shows what simulation it belongs to, and the color of each point indicates whether the debris of that merger event is phase-mixed (blue), tidally disrupted into a stream (red), or still an intact dwarf (green) at the present day. This shows that although mergers that are phase-mixed by the present day typically cross their host's virial radius at early times, there are massive structures that are phase-mixed at present day that cross their host's virial radius within the last 6 Gyr. Note that there is a substantial delay between when the progenitor dwarf galaxy of a radial merger event crosses the virial radius and when it collides with its host galaxy's disk; for example, the dwarf galaxy represented by the blue circle at the top of this figure crossed its host galaxy's virial radius at a lookback time of 6 Gyr, and collided with its host galaxy's disk at a lookback time of 3 Gyr. This figure is a partial reproduction of the top panel of \cite{Horta2023} Figure 2.}\label{fig:mergers}
\end{figure}

It is worth exploring whether recent radial merger events actually occur in cosmological simulations, and what these events look like. \cite{Horta2023} provided a thorough analysis of all massive merger events in the Latte cosmological simulation halos; Figure \ref{fig:mergers} shows their data for the stellar mass of the progenitor of each merger event when it crossed its host galaxy's virial radius, as well as the time when it crossed the virial radius, and a breakdown of the morphology of the debris from each merger event at the present day.

There is a clear trend in the progenitors of the simulated mergers that are phase-mixed today (shown in blue in Figure \ref{fig:mergers}); the progenitors that fall in later are on average more massive than the progenitors that fall in at earlier times. Additionally, we note a distinct gap in intermediate mass mergers ($\log_{10} M_*/M_\odot \sim 8.5$) within the last 6 Gyr of the simulation. This is problematic, because this is where we expect the progenitor of the local caustics to be located based on their progenitor only contributing up to $\sim$20\% of the local stellar halo stars and its merger debris being dynamically young. 

It is possible that this gap is due to the FIRE simulations generating systematically more massive dwarfs compared to other cosmological simulations \citep[e.g. in EAGLE simulations, ][]{Mackereth2018}, which has been pointed out by \cite{Lane2023}. Alternatively, we can flip this problem on its head; the fact that we observe a recent radial merger with intermediate mass in the MW provides a good constraint for future simulations. Are these types of mergers actually rare, or are simulations with different physics able to produce a larger number of intermediate-mass radial merger events at late times?

\subsection{Choice of Recent Merger}

We chose the merger event represented by the blue circle at the top of Figure \ref{fig:mergers} to be our recent radial merger event for comparison with the observed data, as well as the ancient radial merger event. The mass of this merger event is much larger than expected for the VRM or the GSE, and is about half the mass of the LMC. This will increase the rate at which phase-mixing occurs in its host galaxy by substantial introducing disequilibrium into its host galaxy \citep{Donlon2020}. We also expect a more massive progenitor to experience more dynamical friction than a less massive progenitor, which could radialize the orbit of the massive progenitor more than would be expected for a less massive progenitor. However, the large mass of this recent radial merger has one major advantage: because it has a large number of star particles, we do not need to upsample this merger event like we did the ancient merger event. 


We computed the causticality of this radial merger event in an identical way to the ancient merger event, using 8 mock solar regions. The causticality of the recent merger event is shown in Figure \ref{fig:causticality} compared to the causticalities of the ancient merger event and the observed data. 

The causticality of the recent merger event drops much more rapidly than the causticality of the ancient merger event; the causticality of the recent radial merger event agrees with the causticality of the observed data a little over 0.5 Gyr since the time of collision (however, in Appendix \ref{app:bg} we show that this is probably an underestimate of the collision time, especially because of the large mass of this merger's progenitor). This implies that the progenitor of the local caustics could have collided with the MW disk as recently as within the last Gyr. 

While the very massive progenitor of the recent radial merger may artificially accelerate the phase mixing of its debris for the reasons listed above, the increased rate of phase-mixing for the recent radial merger is consistent with the ideas from Section \ref{sec:phase_mixing_rates} that a more recent radial merger will phase mix more quickly.

\section{Discussion} \label{sec:discussion}

\subsection{Regarding the Semi-Analytical Model} 

The semi-analytical model of Section \ref{sec:semi-analytical} indicated that the progenitor of the local phase space folds collided with the MW disk roughly 1-2 Gyr ago. Note that the true collision time can be longer ago than 1--2 Gyr if we are failing to resolve the true number of caustics in the Solar neighborhood, either due to limitations on completeness and/or distance and velocity determination of stars in the merger debris, or because one or more caustics has intrinsically low density. Additionally, it is possible that some caustics are not present in the Solar neighborhood due to the MW having a complicated gravitational potential, which could substantially alter the orbits of these structures. This is especially problematic because the Solar neighborhood is a relatively small region of the Galaxy, so a small change to the orbit of a structure could significantly alter its appearance near the Sun. However, we would need to be underestimating the number of caustics in the Solar neighborhood by a factor of 3 or more in order to obtain a collision time that is consistent with the proposed collision time of the GSE, which is unlikely to be the case.

Additionally, it is possible that one or more of the identified caustics are spuriously generated due to noise, or interloping material from a different physical process such as an unrelated merger, or an in-situ process like the stellar bar, which \cite{Dillamore2024} have proposed as an explanation for the local phase space folds. If this is the case, then the actual number of caustics in the observed region would be lower, and the estimated time of collision would be correspondingly more recent.

In Section \ref{sec:semi-analytical} we assumed that the radial merger debris began on the opposite side of the Galaxy from the Sun (i.e. the Sun is on the right side of the top panel of Figure \ref{fig:semi-analytical}). However, if the progenitor fell in from the same side as the Sun, or in other words, the Sun was located on the left side of the figure instead, then $n_\textrm{wrap}$ would be off by one caustic. In this case, the estimated collision times for the two studies become marginally larger at 2100 Myr and 960 Myr, respectively.

\subsection{On Comparing Observed vs. Simulated Structure} 

We strongly emphasize that it is not enough to observe chevrons in simulated local (or global) phase-space distributions, and then claim that the simulated data matches the observed phase-space distribution simply because chevrons are present. Thick chevrons are present in merger debris long after the progenitor collides with its host (e.g. the botom panel of Figure \ref{fig:simulation}). The observed phase-space distribution contains caustics with $v_s>0$, which indicates that the constituent stars are energy-sorted. This is not true for chevrons present in simulated merger debris long after collision, which appear to have $v_s=0$. These structures are fundamentally different (see Figure \ref{fig:phase_mixing}); chevrons with $v_s=0$ are largely phase-mixed, whereas chevrons with $v_s>0$ are not. In order to make a compelling comparison between the observed structure and simulated structure, the simulated structure must be asymmetric about $v_r=0$, i.e. it must have large causticality.

It is also important to remember that the mixing timescale for radial merger debris is a function of energy with respect to the host galaxy (or equivalently apocenter radius). Another shell is generated once every crossing time \citep{Donlon2020}, and crossing time depends on the distance of a structure's apocenter from the host galaxy. This means that shells at large radii ($\gtrsim45$ kpc) will remain unmixed for long periods of time \citep[$>8$ Gyr at least, as indicated by][]{Han2022b}, whereas caustics in the inner galaxy ($r<45$ kpc) will become noticeably phase-mixed within several Gyr. If the halo is not phase-mixed at large radii, it can still be phase-mixed at small radii, where most of the stars are located; existing shell or cloud structure at large radii in simulations does not indicate the level of phase-mixing in the inner Galaxy.

The phase-space structure of the in-situ halo is shown in the bottom left panel of Figure 8 of \cite{Belokurov2023}. Curiously, the in-situ halo material displays a prominent phase-space fold. We point out that this fold has $v_s=0$, whereas the phase-space folds in the lower-[Fe/H] data in the same figure appear to have $v_s>0$; this suggests that the in-situ halo has been phase-mixing for substantially longer than the debris of the ``last major merger,'' which still has $v_s>0$. Because of this apparent difference in mixing times, it seems likely that the in-situ halo structure was in place before the progenitor of the ``last major merger'' collided with the MW. This is consistent with the in-situ halo either being generated by purely secular processes \citep[e.g.][]{Amarante2020,Yu2020} and/or being caused by a separate merger at early times that is not associated with the ``last major merger'' \citep[e.g.][]{Kruijssen2020,Horta2021,Donlon2022,Malhan2022,Donlon2023}. 

\subsection{Effects of Disequilibrium}

Recent work indicates that it is possible to generate phase-space folds similar to those observed in the \textit{Gaia} data via interactions with a bar potential \citep{Dillamore2023}; this is because stars become trapped within resonances occurring at specific orbital radii, causing alternating excesses and depletions in $r$-$v_r$ phase space. However, it is important to note that the structures generated by resonance with the bar are not energy-sorted, and so they will appear to have $v_s=0$; this is not the case for the caustics in the observed local phase space data, which all have $v_s>0$. As a result, the bar-interaction model of generating phase-space folds does not appear to be applicable to the observed local phase space distribution.

Additionally, there is evidence that the presence of a bar accelerates the phase-mixing process \citep{Davies2023b}. While the m12f FIRE-2 Latte simulation does have a bar, it is relatively short and not long-lived \citep{Ansar2023}; the MW has a much larger bar. If the presence of a bar accelerates phase-mixing, then the MW's larger bar could speed up phase-mixing even further than in our simulated data, which would make it even more unlikely that GSE debris would remain relatively unmixed until the present day. \cite{Davies2023b} claim that they are able to recover the global phase space folds 10 Gyr after the beginning of their simulation; however, these chevrons all have $v_s=0$, which makes them inconsistent with the observed phase-space folds, as their simulated chevrons would have low causticality.

We expect that any disequilibrium in the host galaxy potential will also increase the rate of phase mixing in that galaxy's stellar halo. If there was a lot of disequilibrium at early times and a lot less at late times, this could also influence the accuracy of our collision time estimate. This will depend on the specifics of what disequilibrium is present in the simulation and the MW. We know that the MW is currently in substantial disequilibrium, which could lead to further bias towards older times in our estimated time of collision.

The host halo in the ancient merger simulation experiences two very massive mergers after the infall of the ancient merger; it is not expected that the MW would have corresponding major mergers at this time, which could cause an increase in the rate of phase mixing in the recent merger debris compared to the true radial merger debris in the MW. Additionally, it is possible that our interpolated potential model does not contain some of the transient features excited in the potential of this simulation (although this is unlikely, as there is an excellent match between orbits integrated in the true potential vs. the interpolated potential approximation). These effects would lead to the m12f recent merger debris to be biased low compared to that of the MW, making its predicted collision time too recent. However, it is unclear to what extent major mergers will actually perturb existing radial merger debris, and this will need to be studied more carefully before a quantitative assessment of this bias can be made. It should be noted that these major mergers in m12f cross the host galaxy's virial radius over 4 Gyr after the infall of the recent radial merger, making it unlikely that they substantially affect the phase mixing rates of the simulation until $\sim$4 Gyr after the time of collision.

\section{Conclusions} \label{sec:conclusion}

A substantial amount of previous literature and evidence supports the idea that the MW experienced a merger with a massive dwarf galaxy 8-11 Gyr ago. We provide evidence that the observed local phase space distribution contains caustics that make up $\sim$20\% of the local stellar halo, and can be explained by a dwarf galaxy colliding with the MW disk as recently as within the last Gyr. The observed phase space distribution simultaneously contains a different structure that is dynamically old, and was probably created by a radial merger event at earlier times.

We derived a simple semi-analytical phase mixing model from first principles, which allows us to use the number of caustics within a given range of Galactocentric radii to constrain the collision time of a merger event. Using observed data from \cite{Belokurov2023} and \cite{Wu2023}, we estimate the time of collision of the MW's last major merger to be 1--2 Gyr ago; consistent with predictions of the time that the VRM's impacted the MW disk, but much more recent than the proposed collision time of the GSE. In order for the data to be consistent with the collision time of the GSE, the number of observed caustics would need to be increased by at least a factor of 3.

We analyzed \textit{Gaia} DR3 RVS data within 5 kpc of the Sun. This data contains numerous structures apparent in $r$-$v_r$ phase space, known as phase space folds, caustics, shells, or chevrons (we point out that these terms cannot be used interchangeably). We identify phase space folds in both prograde and retrograde material, similar to the results of \cite{Belokurov2023}. However, the phase space folds in the prograde material are dynamically distinct from the phase space fold in the retrograde material; the phase space folds in the prograde material are asymmetric about the $v_r=0$ line (they are caustic structures), while the phase space fold in the retrograde material is symmetric around the $v_r=0$ line (it is not a caustic structure).

We point out that previous assessments of the local phase space folds has not taken into account that true caustics (which are energy-sorted) and thick chevrons (which are not energy-sorted) are fundamentally different; any model that only generates chevrons in the Solar neighborhood is inconsistent with the observed data. Reports that phase space folds can be recovered $\geq$8 Gyr after collision should be viewed cautiously; chevrons that turn around at $v_r=0$ are not the same dynamical structure as the observed phase space folds, which turn around at $v_r>0$.

The simple phase-mixing model does not include various interactions and time-dependent effects that could alter the rate at which merger debris phase mixes. In order to evaluate how phase-mixed the observed phase space distribution is in a way that accounts for these complicated effects, we upsample a simulation of a GSE analog merger event in the m12f Latte FIRE-2 cosmological zoom-in simulation, and subsequently develop a metric of phase mixing in $r$-$v_r$ phase space. This metric is a 2-dimensional extension of the 1-dimensional phase-mixing metric developed in \cite{Donlon2020}, called ``causticality.'' 

We evaluate the causticality of the observed and simulated data, and find that the observed data and simulated data match best 1.5 Gyr after the dwarf galaxy collides with its host galaxy disk, although it is reasonable that the observed data matches the simulated data until 3 Gyr after collision. At times later than 3 Gyr, the simulated data is significantly more phase-mixed than the observed data; the observed data is certainly not as phase-mixed as one would expect for a merger that collided with the MW disk 8-11 Gyr ago. Including a large bar or other disequilibrium features only reduces the inferred age of the merger. 

Additionally, the simulated merger formed a smooth and monotonically decreasing global radial density profile within 1.3 Gyr after collision, which indicates that the observed MW stellar halo density profile does not rule out the suggestion that a large component of the inner stellar halo was formed from a dwarf galaxy colliding with the MW disk on a radial orbit 1.5--3 Gyr ago.

We also consider a more recent radial merger event in the Latte simulations, and find that it has a causticality consistent with that of the observed data only $\sim$0.5 Gyr after collision. Such a recent collision between a dwarf galaxy and the MW disk might cause disequilibrium that we can observe at the present day, such as the phase space spiral \citep{Antoja2018,Laporte2019,Hunt2021,Hunt2022}.

From our results, we find it highly unlikely that the ``last major merger'' of the MW collided with the MW disk 8-11 Gyr ago; instead, our analysis is consistent with the collision occurring within the last few Gyr. The phase-mixing timescales that arise from the dynamics of stars in the MW are inconsistent with the proposed collision time of the GSE; no known cosmological effects can account for this discrepancy. Consequently, the folds in the local phase space distribution are consistent with being caused by the Virgo Radial Merger \citep{Donlon2019}, which is thought to arise from an [Fe/H]-rich progenitor \citep{Donlon2022,Donlon2023} that recently collided with the MW disk \citep{Donlon2020}. We do not rule out the possibility of one or more early major mergers in addition to the VRM, as has been proposed by several other works \citep{Kruijssen2020,Horta2021,Donlon2022,Donlon2023}. 

As an example, \cite{Donlon2022} and \cite{Donlon2023} provide evidence for an ancient massive, gas-rich merger that was probably quenched upon infall, and is chemodynamically distinct from the stars that make up the \textit{Gaia} Sausage velocity structure. This provides a compelling scenario in which one or more ancient gas-rich mergers occurred in the MW, but a much more recent radial merger event (the VRM) generated the substructure that is typically attributed to the GSE, including the phase space folds analyzed here. 

\acknowledgments

This work was supported by the NASA/NY Space Grant, as well as contributions made by Manit Limlamai and Jake Weiss.\\ 

This work has made use of data from the European Space Agency (ESA) mission {\it Gaia} (\url{https://www.cosmos.esa.int/gaia}), processed by the {\it Gaia} Data Processing and Analysis Consortium (DPAC, \url{https://www.cosmos.esa.int/web/gaia/dpac/consortium}). Funding for the DPAC has been provided by national institutions, in particular the institutions participating in the {\it Gaia} Multilateral Agreement.

This project was made possible by the computing cluster resources provided by the Flatiron Institute Center for Computational Astrophysics.

\software{Numpy \citep{numpy}, Scipy \citep{scipy}, Astropy \citep{astropy}, Matplotlib \citep{matplotlib}, Gizmo \citep{gizmo}, Agama \citep{agama}, Gala \citep{Gala}, Mwahpy \citep{mwahpy}}

\section*{Data Availability}

\textit{Gaia} DR3 data is publicly available at \url{https://gea.esac.esa.int/archive/}. The FIRE data releases are publicly available \citep{Wetzel2023} at \url{http://flathub.flatironinstitute.org/fire}. Additional FIRE simulation data is available at \url{https://fire.northwestern.edu/data}. A public version of the GIZMO code is available at \url{http://www.tapir.caltech.edu/~phopkins/Site/GIZMO.html}.

\bibliographystyle{aasjournal}
\bibliography{main.bib}

%
%
%

\appendix 
\section{Uncertainty in Causticality} \label{app:sigma_c}

Here we compute the uncertainty in Causticality as defined in Equation \ref{eq:c}. We define the quantity \begin{equation}
    \xi_i^\pm = (n_{i,v^+} \pm n_{i,v^-})^2
\end{equation} such that Equation \ref{eq:c} becomes \begin{equation}
    C = \frac{\sum_i^{n_b} \xi^-_i}{\sum_i^{n_b} \xi^+_i}.
\end{equation} Then if we assume that the uncertainty in each bin is Poissonian, i.e. $\sigma_{n_i} = \sqrt{n_i}$, we can compute the uncertainty in $\xi_i^\pm$ using the standard propagation of uncertainty formula: \begin{equation}
    \sigma_{\xi_i^\pm}^2 = 2 \xi^\pm_i \sqrt{\xi^+_i},
\end{equation} and the uncertainty in $C$ follows: \begin{equation} \label{eq:c_unc}
    \sigma_C^2 = \sum_i^{n_b} \frac{1}{(\sum_j^{n_b} \xi^+_j)^2} \sigma_{\xi_i^-}^2 + \sum_i^{n_b} \frac{(\sum_j^{n_b} \xi^-_j)^2}{(\sum_j^{n_b} \xi^+_j)^4}\sigma_{\xi_i^+}^2.
\end{equation} 

Consider the case where there is a background in the data; \begin{equation}
    n'_i = n_i + b_i,
\end{equation} where $n'_i$ is the measured value, $b_i$ is the background, and $n_i$ is the desired signal in a given bin. The uncertainty in $n_i$ is now \begin{equation}
    \sigma^2_{n_i} = \sigma_{n'_i}^2 + \sigma_{b_i}^2.
\end{equation} $\sigma_{n'}^2 = n'$ is Poissonian, but the uncertainty in the background is determined from the Gaussian convolution; \begin{equation}
b_i = \sum_j^{n_b} s_{i,j} n'_{i,j},
\end{equation} where $s_{i,j}$ is the coefficient of each bin's contribution to the convolution, and \begin{equation} \label{eq:app1}
\sigma_{b_i}^2 = \sum_j^{n_b} s^2_{i,j} \sigma_{n'_{i,j}}^2 = \sum_j^{n_b} s^2_{i,j} n'_{i,j}.
\end{equation} In practice, $s_{i,j}\ll1$ (in this work, the maximum $s_{i,j}$ value used is about 0.01), so $\sigma^2_{b_i} \ll \sigma^2_{n'_i}$. Therefore,\begin{equation}
    \sigma^2_{n_i} \approx \sigma_{n'_i}^2,
\end{equation} and Equation \ref{eq:c_unc} holds for the observed data (as long as $n'_i$ is used instead of $n_i$).

\section{Causticality Saturation} \label{app:saturation}

\begin{figure}%
\centering
\includegraphics[width=0.5\textwidth]{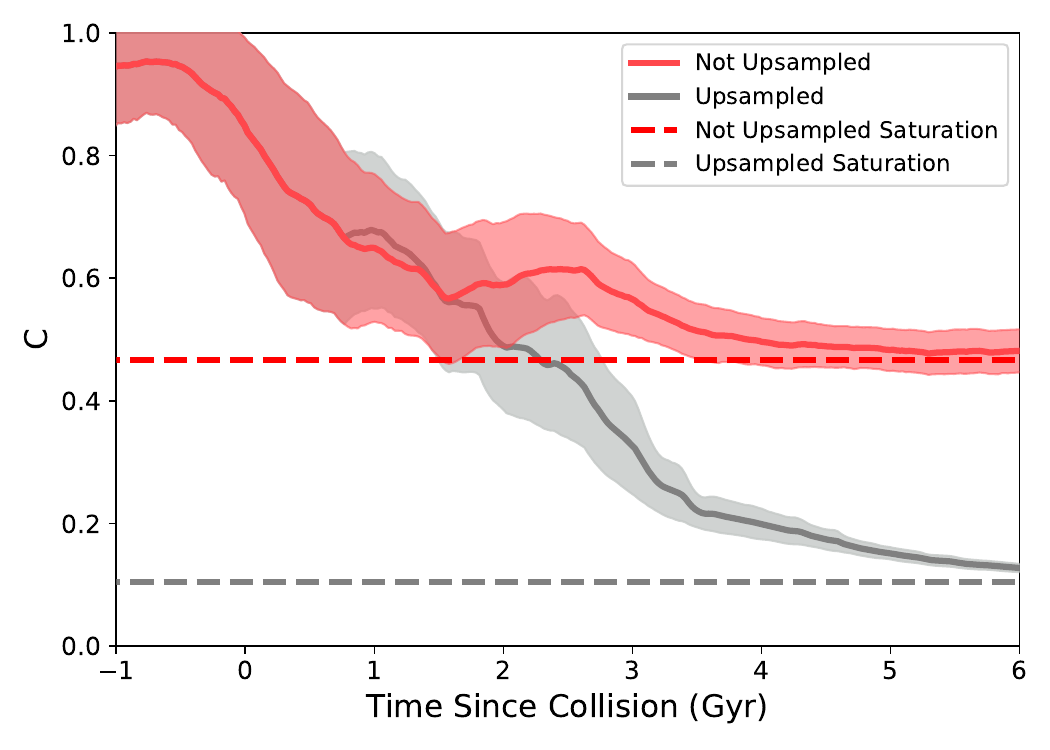}
\caption{Measured causticality for the upsampled simulation (gray) and the not-upsampled simulation (red) of the same ancient radial merger event. The causticality of the debris asymptotically approaches the saturation point towards the end of both simulations. Roughly 2 Gyr after collision the causticality of the upsampled data drops below the saturation point of the not upsampled simulation, but the causticality of the not-upsampled simulation cannot drop below this threshold and ceases to be physically meaningful, illustrating the requirement for the upsampled simulation. }\label{fig:saturation}
\end{figure}

If causticality is computed using a sparse dataset, the computed causticality will be artificially high. This effect is a result of noise in the binned distribution, which generates a strictly positive contribution to the observed causticality. When the distribution phase-mixes to the point where its causticality is indistinguishable from the causticality generated by noise, we consider the causticality of the simulation to be ``saturated''. 

This can be seen in Figure \ref{fig:saturation}; the simulation that was not upsampled maintains substantially larger values of causticality compared to the upsampled simulation at late times.

The saturation point ($S$), which can be interpreted as the lowest possible meaningful causticality measurement for a given distribution given its level of noise, is analytically calculable. We begin with the definition of causticality for a phase-mixed distribution, where the number of particles in each bin is described by a normally distributed random variable $\mathcal{N}$ with mean $n$ and standard deviation $\sigma_n$: \begin{equation}
    S = \frac{\sum_i^{n_b} \left(\mathcal{N} - \mathcal{N}\right)^2}{\sum_i^{n_b} \left(\mathcal{N} + \mathcal{N}\right)^2}.
\end{equation} Here, $n_b$ is the number of bins, although no terms depend on the choice of bin. This allows us to compute $S$ as an expression of the mean sum of square sums and differences of each bin: \begin{equation}
    S = \frac{E\left[\left(\mathcal{N} - \mathcal{N}\right)^2\right]}{E\left[\left(\mathcal{N} + \mathcal{N}\right)^2\right]}.
\end{equation}

The sum or difference of two normally-distributed random variables $\mathcal{N}_i = \mathcal{N}(\mu_i,\sigma_i)$ is itself a normally-distributed random variable: \begin{equation}
    \mathcal{N}_\pm = \mathcal{N}_1 \pm \mathcal{N}_2 = \mathcal{N}(\mu_1 \pm \mu_2,\sigma_1^2 + \sigma_2^2).
\end{equation} This, in combination with the definition of variance, \begin{equation}
    Var(\mathcal{N}) = E\left[\mathcal{N}^2\right] - E\left[\mathcal{N}\right]^2,
\end{equation} can then be used to obtain: \begin{equation}
    S = \frac{E\left[\mathcal{N}_-^2\right]}{E\left[\mathcal{N}_+^2\right]} = \frac{2\sigma_n^2}{2\sigma_n^2 + (2n)^2}.
\end{equation}

If we assume that the uncertainty in the number of particles in each bin is described by a Poisson distribution, so that $\sigma_n^2=n$, this reduces to \begin{equation}\label{eq:S}
    S = \frac{1}{2n+1}.
\end{equation}

The analytically computed saturation points for the upsampled and non-upsampled simulations are shown as dashed lines in Figure \ref{fig:saturation}, and agree well with the minimum values of causticality for each simulation. If one wants to observe a causticality below the saturation point of a given simulation, then that simulation will need to be upsampled or re-simulated with higher resolution.

The saturation point of the observed data is roughly 0.1, so we are not concerned that the measured causticality of the observed data is saturated.


\section{The Effect of Bin Size on Causticality} \label{app:binning}

\begin{figure}%
\centering
\includegraphics[width=0.5\textwidth]{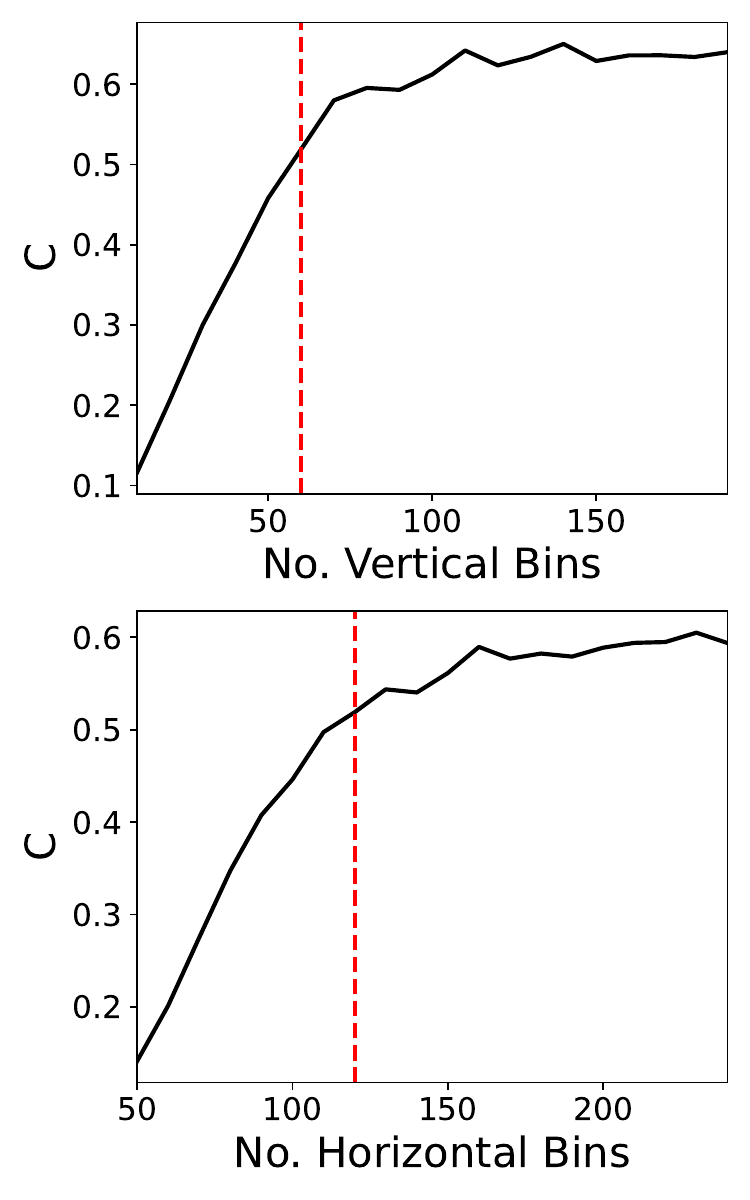}
\caption{Causticality of the observed data for different binning schemes. The top panel shows how causticality changes as one changes the number of bins in $v_r$; the bottom panel shows the variation of causticality based on the number of $r$ bins. For a small number of bins, causticality drops to 0 because small-scale features in the data are averaged out. For a large number of bins, causticality is increased because small-scale features are more apparent in the residual. The choice of bin size in this work is shown as a dashed red line.}\label{fig:c_bins}
\end{figure}

Causticality is a discrete statistic, i.e. the data must be binned in $r$ and $v_r$ in order to compute $C$. As such, it is expected that the choice of bin size can affect the measured $C$ value for a given dataset. 

Figure \ref{fig:c_bins} shows the dependence of $C$ on the size of bins in $r$ and $v_r$. In general, as bins are made larger, causticality decreases; as bins are made smaller, causticality increases. This is because small-scale structure is more prominent in smaller bin sizes, and is averaged out over large bin sizes. As a result, when folding the residual, it is more likely that any asymmetric features will be removed if the bin size is larger. 

We do not want our data to be dominated by noise (if the bin size is too small), but we also do not want to average over interesting substructure (if the bin size is too big). It is reasonable to place the choice of bin size at roughly the ``knee'' in Figure \ref{fig:c_bins}, which should balance the bin size between too-big and too-small, as is the case for our main study.

\section{The Effect of Background on Causticality} \label{app:bg}

\begin{figure}%
\centering
\includegraphics[width=0.5\textwidth]{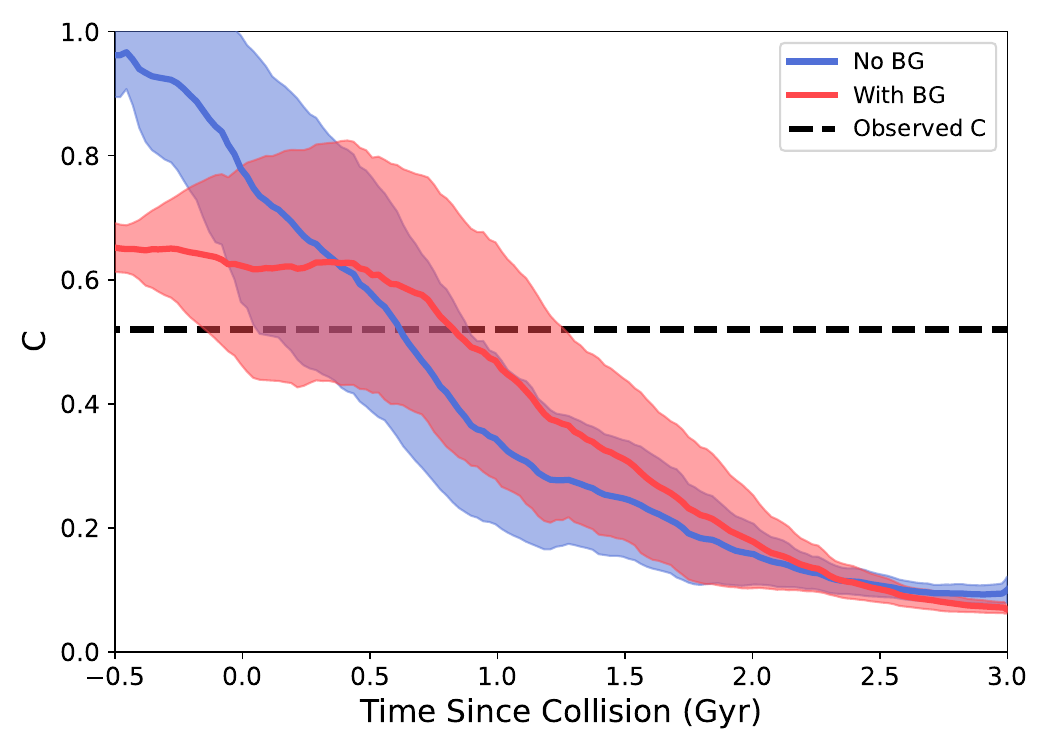}
\caption{Causticality vs time for the recent merger simulation without any background (blue) and with the addition of a background consisting of particles from other accretion events (red). The causticality of the observed data is shown with a dashed black line. Until $\sim$0.5 Gyr after collision, the background dominates the mock Solar circle and causes causticality to drop compared to the causticality of the isolated merger debris; afterwards, the presence of a background causes a slightly increase in causticality compared to the isolated merger debris, which causes our estimate of collision time in Section \ref{sec:recent_rme} to be a slight underestimate. }\label{fig:c_bg}
\end{figure}

We subtracted a background model from the observed phase space data in Section \ref{sec:data} in order to isolate phase space folds. However, we expect that there will still be some background present in the phase space distribution of the data minus the background. The presence of background in the residual could impact the measured value of causticality for the observed data; here, we illustrate how the addition of a background impacts causticality in a radial merger simulation.

Figure \ref{fig:c_bg} shows a comparison of causticality for the recent merger simulation and the same data plus the addition of a background. The background consisted of all particles identified as belonging to major mergers by \cite{Horta2023}, excluding the merger we are interested in. The background data was also subject to the same angular momentum cuts as the recent merger particles. 

Before and shortly after the radial merger progenitor collides with the host galaxy disk, the causticality of the simulated data with a background is much lower than the causticality of the simulated data with no background; this is because the merger debris does not contribute many particles to the mock Solar circle until $\sim$0.5 Gyr after collision, so the measured causticality is diluted by the background accreted material. 

After $\sim$0.5 Gyr since collision, when the recent merger event begins to contribute a substantial number of particles to the mock Solar circle, the presence of a background drives the causticality up by a small amount compared to the causticality of the isolated merger debris. We expect that the presence of background in the observed data causes causticality to be a slight underestimate of the actual time of collision, because any background in the observed phase space residual will drive the measured causticality up by a small amount. Based on the example shown in Figure \ref{fig:c_bg}, this effect is expected to adjust the predicted time of collision by a few tenths of a Gyr.

The small discrepancy between the causticality of the isolated merger debris and the merger debris plus background at late times ($>2.5$ Gyr) is because the presence of additional merger debris lowers the saturation point of the simulated data, as the average occupation of each bin is slightly larger for the merger debris plus background than it is for just the merger debris. Additionally, the uncertainty in the causticality of the merger debris plus a background is slightly larger than the uncertainty in causticality for the isolated merger debris.

Originally, we attempted to include all particles from the simulation (which includes disk particles, in-situ halo particles, accreted particles, etc.) as background. However, in this simulation the merger that we care about has a mean $L_z$ that is roughly 1000 kpc km s\inv from the mean disk $L_z$, compared to a separation of about 2500 $L_z$ between the MW disk and the accreted stars in the local Solar region. This smaller separation in $L_z$ meant that it was not possible to separate in-situ material from the accreted material as easily in the simulation just based on kinematics, and as a result the disk star ``background'' contributed many times more particles to each mock Solar region than the actual merger event we are interested in.

\end{document}